\documentclass[pra,aps,showpacs,epsf,floats]{revtex4}
\usepackage{amssymb}
\usepackage{amsfonts}
\usepackage{amsmath}

\usepackage{graphicx}

\begin{document}

\title{Quantum Stabilizer Codes for Correlated and Asymmetric Depolarizing
Errors}
\author{Carlo Cafaro$^{1}$ and Stefano Mancini$^{2}$}
\affiliation{$^{1,2}$School of Science and Technology, Physics Division, University of
Camerino, I-62032 Camerino, Italy}

\begin{abstract}
We study the performance of common quantum stabilizer codes in the presence
of asymmetric and correlated errors. Specifically, we consider the
depolarizing noisy quantum memory channel and perform quantum error
correction via the five and seven-qubit stabilizer codes. We characterize
these codes by means of the entanglement fidelity as function of the error
probability and the degree of memory. We show that their performances are
lowered by the presence of correlations and we compute the error probability
threshold values for codes effectiveness. Furthermore, we uncover that the
asymmetry in the error probabilities does not affect the performance of the
five-qubit code while it does affect the performance of the seven-qubit code
which results less effective when considering correlated and symmetric
depolarizing errors but more effective for correlated and asymmetric errors.
\end{abstract}

\maketitle

\section{Introduction}

The most important obstacle in quantum information processing is
decoherence. It causes a quantum computer to lose its quantum properties
destroying its performance advantages over a classical computer. The
unavoidable interaction between the open quantum processor and its
environment corrupts the information stored in the system and causes errors
that may lead to wrong outputs. In general, environments may be very complex
systems characterized by many uncontrollable degrees of freedom. A useful
active strategy to defend quantum coherence of a processing against
environmental noise is that of quantum error correcting codes (QECC) \cite%
{perimeter, knill97, calderbank97} where, in analogy to classical
information theory, quantum information is stabilized by using redundant
encoding and measurements.

The formal mathematical description of the qubit-environment interaction is
often given in terms of quantum channels. Quantum error correction is
usually developed under the assumption of i.i.d. (identically and
independently distributed) errors. These error models are characterized by
memoryless communication channels $\Lambda $ such that $n$-channel uses is
given by $\Lambda ^{\left( n\right) }=\Lambda ^{\otimes n}$. In such cases
of complete independent decoherence, qubits interact with their own
environments which do not interact with each other. However, in actual
physical situations, qubits may interact with a common environment which
unavoidably introduces correlations in the noise. For instance, there are
situations where qubits in a ion trap set-up are collectively coupled to
their vibrational modes \cite{garg96}. In other situations, different qubits
in a quantum dot design are coupled to the same lattice, thus interacting
with a common thermal bath of phonons \cite{loss98}. The exchange of bosons
between qubits causes spatial and temporal correlations that violate the
condition of error independence \cite{hwang01}. Memory effects introduce
correlations among channel uses with the consequence that $\Lambda ^{\left(
n\right) }\neq \Lambda ^{\otimes n}$. Recent studies try to characterize the
effect of correlations on the performance of QECCs \cite{clemens04,
klesse05, shabani08, d'arrigo08, carlo-PLA}. It appears that correlations
may have negative \cite{klesse05} or positive \cite{shabani08} impact on
QECCs depending on the features of the error model being considered.

Furthermore, the noise may be asymmetric. Most of the quantum computing
devices \cite{astafiev04} are characterized by relaxation times ($\tau _{%
\text{relaxation}}$) that are one-two orders of magnitude larger than the
corresponding dephasing times ($\tau _{\text{dephasing}}$). Relaxation leads
to both bit-flip and phase-flip errors, whereas dephasing (loss of phase
coherence, phase-shifting) only leads to phase-flip errors. Such asymmetry
between $\tau _{\text{relaxation}}$ and $\tau _{\text{dephasing}}$
translates to an asymmetry in the occurrence probability of bit-flip ($p_{X}$%
) and phase-flip errors ($p_{Z}$). The ratio $\frac{p_{Z}}{p_{X}}$ is known
as the channel asymmetry. Quantum error correction schemes should be
designed in such a way that no resources (time and qubits) are wasted in
attempting to detect and correct errors that may be relatively unlikely to
occur. Quantum codes should be designed in order to exploit this asymmetry
and provide better performance by neglecting the correction of less probable
errors \cite{ioffe07, evans07, stephens08}. Indeed, examples of efficient
quantum error-correcting codes (for instance, asymmetric stabilizer CSS\
codes) taking advantage of this asymmetry are given by families of codes of
the Calderbank-Shor-Steane (CSS) type \cite{sarvepalli08, aly08}.

Following these lines of investigations, in this article we study the
performance of common quantum stabilizer codes in the presence of asymmetric
and correlated errors. Specifically, we consider the depolarizing noisy
quantum memory channel and perform quantum error correction via the five and
seven-qubit stabilizer codes \cite{gottesman97}. We characterize the
performance of the codes by means of the entanglement fidelity $\mathcal{F}%
\left( \mu \text{, }p\right) $ \cite{schumacher96} as function of the error
probability $p$ and degree of memory $\mu $ (correlations). We show that the
performance of both codes is lowered in the presence of correlations and
error probability threshold values for code effectiveness are computed vs.
the degree of memory $\mu $. The error correction schemes here considered
only work for low values of $\mu $. Furthermore, we uncover that the
asymmetry in the error probabilities does not affect the performance of the
five-qubit code while it does affect the performance of the seven-qubit code
which results less effective when considering correlated and symmetric
depolarizing errors, but more effective for correlated and asymmetric errors.

The layout of the paper is as follows. In Section II, we consider a
depolarizing noisy quantum memory channel characterized by symmetric error
probabilities and QEC is performed via the $\left[ \left[ 5\text{, }1\text{, 
}3\right] \right] $ stabilizer code. The performance of quantum error
correcting codes is quantified by means of the entanglement fidelity $%
\mathcal{F}^{\left[ \left[ 5,1,3\right] \right] }\left( \mu \text{, }%
p\right) $ as function of the error probability $p$ and degree of memory $%
\mu $. In Section III, QEC is performed via the $\left[ \left[ 7\text{, }1%
\text{, }3\right] \right] $-CSS stabilizer code. The performance of quantum
error correcting codes is quantified by means of the entanglement fidelities 
$\mathcal{F}_{\text{Set-}1}^{\left[ \left[ 7,1,3\right] \right] }\left( \mu 
\text{, }p\right) $ and $\mathcal{F}_{\text{Set-}2}^{\left[ \left[ 7,1,3%
\right] \right] }\left( \mu \text{, }p\right) $ as function of the error
probability $p$ and degree of memory $\mu $ evaluated for two different
allowable sets of correctable error operators. In Section IV, for asymmetric
error probabilities and correlated noise errors, we show that the
seven-qubit code can outperform the five qubit-code and it also endowed with
a better threshold curve $\mu _{\text{threshold}}=\mu _{\text{threshold}%
}\left( p\right) $ where error correction is performed in an effective way.
Finally, in Section V we present our final remarks.

\section{ The Five-Qubit Code: Symmetric Error Probabilities and Correlations%
}

In this Section, we consider a depolarizing noisy quantum memory channel
with symmetric error probabilities and QEC is performed via the $\left[ %
\left[ 5\text{, }1\text{, }3\right] \right] $ stabilizer code. The
performance of quantum error correcting codes is quantified by means of the
entanglement fidelity $\mathcal{F}^{\left[ \left[ 5,1,3\right] \right]
}\left( \mu \text{, }p\right) $ as function of the error probability $p$ and
degree of memory $\mu $.

\emph{Error Model}. The depolarizing channel is especially easy to analyze
in the context of quantum error-correction because it has a simple
interpretation in terms of the four basic errors $I$, $X$, $Y$, $Z$ which
are the most commonly used in the analysis of quantum codes. However, this
error model is rather general since the ability to error-correct the
depolarizing channel automatically implies the ability to error-correct an
arbitrary single qubit quantum operation. To simplify the notation, we may
choose to omit sometimes the symbol of tensor product "$\otimes $" in the
expressions for the error operators of weight greater than one.

Consider five qubits and Markov correlated errors in a depolarizing quantum
channel $\Lambda ^{(5)}(\rho )$, 
\begin{equation}
\Lambda ^{(5)}(\rho )=\sum_{i_{1}\text{, }i_{2}\text{, }i_{3}\text{, }i_{4}%
\text{, }%
i_{5}=0}^{3}p_{i_{5}|i_{4}}p_{i_{4}|i_{3}}p_{i_{3}|i_{2}}p_{i_{2}|i_{1}}p_{i_{1}}%
\left[ A_{i_{5}}A_{i_{4}}A_{i_{3}}A_{i_{2}}A_{i_{1}}\rho A_{i_{1}}^{\dag
}A_{i_{2}}^{\dag }A_{i_{3}}^{\dag }A_{i_{4}}^{\dagger }A_{i_{5}}^{\dagger }%
\right] \text{,}
\end{equation}%
where $A_{0}\equiv I$, $A_{1}\equiv X$, $A_{2}\equiv Y$, $A_{3}\equiv Z$ are
the Pauli operators defined as,%
\begin{equation}
I\left\vert q\right\rangle \overset{\text{def}}{=}\left\vert q\right\rangle 
\text{, }X\left\vert q\right\rangle \overset{\text{def}}{=}\left\vert
q\oplus 1\right\rangle \text{, }Z\left\vert q\right\rangle \overset{\text{def%
}}{=}\left( -1\right) ^{q}\left\vert q\right\rangle \text{, }Y\left\vert
q\right\rangle \overset{\text{def}}{=}i\left( -1\right) ^{q}\left\vert
q\oplus 1\right\rangle \text{,}
\end{equation}%
with $q=0$, $1$ and $X$, $Y$ and $Z$ given by,%
\begin{equation}
X=\left( 
\begin{array}{cc}
0 & 1 \\ 
1 & 0%
\end{array}%
\right) \text{, }Y=iXZ=\left( 
\begin{array}{cc}
0 & -i \\ 
i & 0%
\end{array}%
\right) \text{, }Z=\left( 
\begin{array}{cc}
1 & 0 \\ 
0 & -1%
\end{array}%
\right) \text{.}
\end{equation}%
The coefficients $p_{i_{l}|i_{m}}$ (conditional probabilities) with $l$, $m$ 
$\in \left\{ 0\text{, }1\text{, .., }5\right\} $ satisfy the normalization
condition,%
\begin{equation}
\sum_{i_{1}\text{, }i_{2}\text{, }i_{3}\text{, }i_{4}\text{, }%
i_{5}=0}^{3}p_{i_{5}|i_{4}}p_{i_{4}|i_{3}}p_{i_{3}|i_{2}}p_{i_{2}|i_{1}}p_{i_{1}}=1%
\text{.}
\end{equation}%
For the depolarizing channel $\Lambda ^{(5)}(\rho )$, coefficients $%
p_{i_{l}|i_{m}}$ are considered as, 
\begin{equation}
p_{k|j}\overset{\text{def}}{=}(1-\mu )p_{k}+\mu \delta _{k\text{, }j}\text{,}%
\quad p_{k=0}=1-p\text{,}\;p_{k=1\text{, }2\text{, }3}=p/3\text{,}
\label{conditional}
\end{equation}%
where $p\in \left[ 0\text{, }1\right] $ denotes the error probability, $\mu
\in \left[ 0\text{, }1\right] $ represents the degree of memory ($\mu =0$
gives the uncorrelated errors and $\mu =1$ gives perfectly correlated
errors) and $p_{k|j}$ is the probability of error $k$ on qubit $j$. To
simplify the notation, we may choose to suppress the bar "$|$" appearing in
the conditional probabilities ($p_{k|j}\equiv p_{kj}$). Furthermore, since
we are initially assuming $p_{1}=p_{2}=p_{3}=p/3$, we are in the case of
symmetric error probabilities.

\emph{Error Operators}. In an explicit way, the depolarizing channel $%
\Lambda ^{(5)}(\rho )$ can be written as,%
\begin{equation}
\Lambda ^{(5)}(\rho )=\sum_{k=0}^{2^{10}-1}A_{k}^{\prime }\rho A_{k}^{\prime
\dagger }\text{,}  \label{nota}
\end{equation}%
where $A_{k}^{\prime }$ are the enlarged error operators acting on the five
qubit quantum states. The cardinality of the error operators defining $%
\Lambda ^{(5)}(\rho )$ is $2^{10}$ and is obtained by noticing that,%
\begin{equation}
\sum_{m=0}^{5}3^{m}\binom{5}{m}=2^{10}\text{,}
\end{equation}%
where $3^{m}\binom{5}{m}$ is the cardinality of weight-$m$ error operators $%
A_{k}^{\prime }$. More details on the explicit expressions for weight-$0$
and weight-$1$ appear in the Appendix A.

\emph{Encoding}. The $\left[ \left[ 5\text{, }1\text{, }3\right] \right] $
code is the smallest single-error correcting quantum code \cite{laflamme96}.
Of all QECCs that encode $1$ qubit of data and correct all single-qubit
errors, the $\left[ \left[ 5\text{, }1\text{, }3\right] \right] $ is the
most efficient, saturating the quantum Hamming bound. It encodes $k=1$ qubit
in $n=5$ qubits. The cardinality of its stabilizer group $\mathcal{S}$ is $%
\left\vert \mathcal{S}\right\vert =2^{n-k}=16$ and the set $\mathcal{B}_{%
\mathcal{S}}^{\left[ \left[ 5,1,3\right] \right] }$ of $n-k=4$ group
generators is given by \cite{nielsen00},%
\begin{equation}
\mathcal{B}_{\mathcal{S}}^{\left[ \left[ 5,1,3\right] \right] }\overset{%
\text{def}}{=}\left\{ X^{1}Z^{2}Z^{3}X^{4}\text{, }X^{2}Z^{3}Z^{4}X^{5}\text{%
, }X^{1}X^{3}Z^{4}Z^{5}\text{, }Z^{1}X^{2}X^{4}Z^{5}\right\} \text{.}
\end{equation}%
The distance of the code is $d=3$ and therefore the weight of the smallest
error $A_{l}^{\prime \dagger }A_{k}^{\prime }$ \ that cannot be detected by
the code is $3$. Finally, we recall that it is a non-degenerate code since
the smallest weight for elements of $\mathcal{S}$ (other than identity) is $%
4 $ and therefore it is greater than the distance $d=3$. The encoding for
the $\left[ \left[ 5\text{, }1\text{, }3\right] \right] $ code is given by 
\cite{laflamme96},%
\begin{equation}
\left\vert 0\right\rangle \rightarrow \left\vert 0_{L}\right\rangle =\frac{1%
}{4}\left[ 
\begin{array}{c}
\left\vert 00000\right\rangle +\left\vert 11000\right\rangle +\left\vert
01100\right\rangle +\left\vert 00110\right\rangle +\left\vert
00011\right\rangle +\left\vert 10001\right\rangle -\left\vert
01010\right\rangle -\left\vert 00101\right\rangle + \\ 
\\ 
-\left\vert 10010\right\rangle -\left\vert 01001\right\rangle -\left\vert
10100\right\rangle -\left\vert 11110\right\rangle -\left\vert
01111\right\rangle -\left\vert 10111\right\rangle -\left\vert
11011\right\rangle -\left\vert 11101\right\rangle%
\end{array}%
\right] \text{,}  \label{code5}
\end{equation}%
and,%
\begin{equation}
\left\vert 1\right\rangle \rightarrow \left\vert 1_{L}\right\rangle =\frac{1%
}{4}\left[ 
\begin{array}{c}
\left\vert 11111\right\rangle +\left\vert 00111\right\rangle +\left\vert
10011\right\rangle +\left\vert 11001\right\rangle +\left\vert
11100\right\rangle +\left\vert 01110\right\rangle -\left\vert
10101\right\rangle -\left\vert 11010\right\rangle + \\ 
\\ 
-\left\vert 01101\right\rangle -\left\vert 10110\right\rangle -\left\vert
01011\right\rangle -\left\vert 00001\right\rangle -\left\vert
10000\right\rangle -\left\vert 01000\right\rangle -\left\vert
00100\right\rangle -\left\vert 00010\right\rangle%
\end{array}%
\right] \text{.}
\end{equation}%
\emph{Recovery Operators}. Recall that any error belonging to the Pauli
group of $n$-qubits, $E\in \mathcal{P}_{n}$, can be written as,%
\begin{equation}
E=i^{\xi }\sigma _{k_{1}}^{1}\otimes \text{...}\otimes \sigma _{k_{n}}^{n}%
\text{,}  \label{EQ1}
\end{equation}%
where $\xi =0$, $1$, $2$, $3$ and the superscripts on the $\sigma
_{k_{l}}^{l}$ label the qubits $l=1$,..., $n$. Furthermore, the subscripts
take values $k_{l}=0$, $x$, $y$, $z$ (therefore, $\sigma _{0}\equiv I$, $%
\sigma _{x}\equiv X$, $\sigma _{y}\equiv Y$, $\sigma _{z}\equiv Z$) and $%
\sigma _{0}^{l}=I^{l}$ is the identity operator on the $l^{th}$ qubit.
Notice that since $\sigma _{y}^{l}=-i\sigma _{x}^{l}\sigma _{z}^{l}$, (\ref%
{EQ1}) can be rewritten as,%
\begin{equation}
E=i^{\xi ^{\prime }}\sigma _{x}\left( a\right) \sigma _{z}\left( b\right) 
\text{,}  \label{EQ2}
\end{equation}%
where $a=a_{1}$...$a_{n}$ and $b=b_{1}$...$b_{n}$ are the bit strings of
length $n$ with,%
\begin{equation}
\sigma _{x}\left( a\right) \equiv \left( \sigma _{x}^{1}\right)
^{a_{1}}\otimes \text{...}\otimes \left( \sigma _{x}^{n}\right) ^{a_{n}}%
\text{ and, }\sigma _{z}\left( b\right) \equiv \left( \sigma _{z}^{1}\right)
^{b_{1}}\otimes \text{...}\otimes \left( \sigma _{z}^{n}\right) ^{b_{n}}%
\text{.}
\end{equation}%
Although the factor $i^{\xi ^{\prime }}$ in (\ref{EQ2}) is needed to insure
that $\mathcal{P}_{n}$ is a group, in many discussions it is only necessary
to work with the quotient group $\mathcal{P}_{n}/\left\{ \pm I\text{, }\pm
iI\right\} $.

There is a $1-1$ correspondence between $\mathcal{P}_{n}/\left\{ \pm I\text{%
, }\pm iI\right\} $ and the $2n$-dimensional binary vector space $F_{2}^{2n}$
whose elements are bit strings of length $2n$ \cite{calderbank98}. A vector $%
v\in F_{2}^{2n}$ is denoted $v=\left( a|b\right) $, where $a=a_{1}$...$a_{n}$
and $b=b_{1}$...$b_{n}$ are bit strings of length $n$. Scalars take values
in the Galois field $F_{2}=\left\{ 0\text{, }1\right\} $ and vector addition
adds components modulo $2$. In short, $E=i^{\xi }\sigma _{x}\left(
a_{l}\right) \sigma _{z}\left( b_{l}\right) \in \mathcal{P}%
_{n}\leftrightarrow v_{l}=\left( a_{l}|b_{l}\right) \in F_{2}^{2n}$. For a
quantum stabilizer code $\mathcal{C}$ with generators $g_{1}$,..., $g_{n-k}$
and parity check matrix $H$, the error syndrome $S(E)$ for an error $E\in 
\mathcal{P}_{n}\leftrightarrow v_{E}=\left( a_{E}|b_{E}\right) \in
F_{2}^{2n} $ is given by the bit string,%
\begin{equation}
S(E)=Hv_{E}=l_{1}\text{...}l_{n-k}\text{,}
\end{equation}%
where,%
\begin{equation}
l_{j}=H^{T}\left( j\right) \cdot v_{E}=\left\langle v_{j}\text{, }%
E\right\rangle \text{,}
\end{equation}%
with $v_{j}=\left( a_{j}|b_{j}\right) $ the image of the generators $g_{j}$
and $\left\langle \cdot \text{, }\cdot \right\rangle $ the symbol for the
symplectic inner product \cite{calderbank98}. Furthermore, recall that
errors with non-vanishing error syndrome are detectable and that a set of
invertible error operators $\mathcal{A}_{\text{correctable }}$ is
correctable if the set given by $\mathcal{A}_{\text{correctable }}^{\dagger }%
\mathcal{A}_{\text{correctable }}$is detectable \cite{knill02}. It is
straightforward, though tedious, to check that (see Appendix A),%
\begin{equation}
S\left( A_{l}^{\prime \dagger }A_{k}^{\prime }\right) \neq 0\text{, with }l%
\text{, }k\in \left\{ 0\text{, }1\text{,..., }15\right\} \text{,}
\end{equation}%
where $S\left( A_{k}^{\prime }\right) $ is the error syndrome of the error
operator $A_{k}^{\prime }$ defined as,%
\begin{equation}
S\left( A_{k}^{\prime }\right) \overset{\text{def}}{=}H^{\left[ \left[ 5,1,3%
\right] \right] }v_{A_{k}^{\prime }}\text{.}
\end{equation}%
The quantity $H^{\left[ \left[ 5,1,3\right] \right] }$ is the check matrix
for the five-qubit code \cite{nielsen00},%
\begin{equation}
H^{\left[ \left[ 5,1,3\right] \right] }\overset{\text{def}}{=}\left( 
\begin{array}{ccccccccccc}
1 & 1 & 0 & 0 & 0 & | & 0 & 0 & 1 & 0 & 1 \\ 
0 & 1 & 1 & 0 & 0 & | & 1 & 0 & 0 & 1 & 0 \\ 
0 & 0 & 1 & 1 & 0 & | & 0 & 1 & 0 & 0 & 1 \\ 
0 & 0 & 0 & 1 & 1 & | & 1 & 0 & 1 & 0 & 0%
\end{array}%
\right) \text{,}  \label{check1}
\end{equation}%
and $v_{A_{k}^{\prime }}$ is the vector in the $10$-dimensional binary
vector space $F_{2}^{10}$ corresponding to the error operator $A_{k}^{\prime
}$. The set of correctable error operators is given by,%
\begin{equation}
\mathcal{A}_{\text{correctable}}=\left\{ A_{0}^{\prime }\text{, }%
A_{1}^{\prime }\text{, }A_{2}^{\prime }\text{, }A_{3}^{\prime }\text{, }%
A_{4}^{\prime }\text{, }A_{5}^{\prime }\text{, }A_{6}^{\prime }\text{, }%
A_{7}^{\prime }\text{, }A_{8}^{\prime }\text{, }A_{9}^{\prime }\text{, }%
A_{10}^{\prime }\text{, }A_{11}^{\prime }\text{, }A_{12}^{\prime }\text{, }%
A_{13}^{\prime }\text{, }A_{14}^{\prime }\text{, }A_{15}^{\prime }\right\}
\subseteq \mathcal{A}\text{,}
\end{equation}%
where the cardinality of $\mathcal{A}$ defining the channel in (\ref{nota})
equals $2^{10}$. All weight zero and one error operators satisfy the error
correction conditions \cite{nielsen00, laflamme07},%
\begin{equation}
\left\langle i_{L}|A_{l}^{\prime \dagger }A_{m}^{\prime }|j_{L}\right\rangle
=\alpha _{lm}\delta _{ij}\text{,}
\end{equation}%
for $l$, $m\in \left\{ 0\text{, }1\text{,..., }15\right\} $ and $i$, $j\in
\left\{ 0\text{, }1\right\} $ with $\left\langle i_{L}|j_{L}\right\rangle
=\delta _{ij}$. The two \ sixteen-dimensional orthogonal subspaces $\mathcal{%
V}^{0_{L}}$ and $\mathcal{V}^{1_{L}}$ of $\mathcal{H}_{2}^{5}$ generated by
the action of $\mathcal{A}_{\text{correctable}}$ on $\left\vert
0_{L}\right\rangle $ and $\left\vert 1_{L}\right\rangle $ are given by,%
\begin{equation}
\mathcal{V}^{0_{L}}=Span\left\{ \left\vert v_{k}^{0_{L}}\right\rangle =\frac{%
A_{k}^{\prime }}{\sqrt{\tilde{p}_{k}}}\left\vert 0_{L}\right\rangle \text{, }%
\right\} \text{,}
\end{equation}%
with $k=0$, $1$,..., $15$ and,%
\begin{equation}
\mathcal{V}^{1_{L}}=Span\left\{ \left\vert v_{k}^{1_{L}}\right\rangle =\frac{%
A_{k}^{\prime }}{\sqrt{\tilde{p}_{k}}}\left\vert 1_{L}\right\rangle \right\} 
\text{,}
\end{equation}%
respectively. Notice that $\left\langle v_{l}^{i_{L}}|v_{l^{\prime
}}^{j_{L}}\right\rangle =\delta _{ll^{\prime }}\delta _{ij}$ with $l$, $%
l^{\prime }\in \left\{ 0\text{, }1\text{,..., }15\right\} $ and $i$, $j\in
\left\{ 0\text{, }1\right\} $. Therefore, it follows that $\mathcal{V}%
^{0_{L}}\oplus \mathcal{V}^{1_{L}}=\mathcal{H}_{2}^{5}$. The recovery
superoperator $\mathcal{R}\leftrightarrow \left\{ R_{l}\right\} $ with $l=1$%
,...,$16$ is defined as \cite{knill97},%
\begin{equation}
R_{l}\overset{\text{def}}{=}V_{l}\sum_{i=0}^{1}\left\vert
v_{l}^{i_{L}}\right\rangle \left\langle v_{l}^{i_{L}}\right\vert \text{,}
\label{recovery}
\end{equation}%
where the unitary operator $V_{l}$ is such that $V_{l}\left\vert
v_{l}^{i_{L}}\right\rangle =\left\vert i_{L}\right\rangle $ for $i\in
\left\{ 0\text{, }1\right\} $. Notice that (see Appendix A for the explicit
expressions of recovery operators),%
\begin{equation}
R_{l}\overset{\text{def}}{=}V_{l}\sum_{i=0}^{1}\left\vert
v_{l}^{i_{L}}\right\rangle \left\langle v_{l}^{i_{L}}\right\vert =\left\vert
0_{L}\right\rangle \left\langle v_{l}^{0_{L}}\right\vert +\left\vert
1_{L}\right\rangle \left\langle v_{l}^{1_{L}}\right\vert \text{.}
\label{may1}
\end{equation}%
Notice that $\mathcal{R}\leftrightarrow \left\{ R_{l}\right\} $ is a trace
preserving quantum operation, $\sum_{l=1}^{16}R_{l}^{\dagger
}R_{l}=I_{32\times 32}$, since $\left\{ \left\vert
v_{l}^{i_{L}}\right\rangle \right\} $ with $l=1$,...,$16$ and $i_{L}\in
\left\{ 0\text{, }1\right\} $ is an orthonormal basis for $\mathcal{H}%
_{2}^{5}$. Finally, the action of this recovery operation $\mathcal{R}$ on
the map $\Lambda ^{\left( 5\right) }\left( \rho \right) $ in (\ref{nota})
yields,%
\begin{equation}
\Lambda _{\text{recover}}^{\left( 5\right) }\left( \rho \right) \equiv
\left( \mathcal{R\circ }\Lambda ^{(5)}\right) \left( \rho \right) \overset{%
\text{def}}{=}\sum_{k=0}^{2^{10}-1}\sum\limits_{l=1}^{16}\left(
R_{l}A_{k}^{\prime }\right) \rho \left( R_{l}A_{k}^{\prime }\right)
^{\dagger }\text{.}  \label{r5}
\end{equation}

\emph{Entanglement Fidelity}. Entanglement fidelity is a useful performance
measure of the efficiency of quantum error correcting codes. It is a
quantity that keeps track of how well the state and entanglement of a
subsystem of a larger system are stored, without requiring the knowledge of
the complete state or dynamics of the larger system. More precisely, the
entanglement fidelity is defined for a mixed state $\rho =\sum_{i}p_{i}\rho
_{i}=$tr$_{\mathcal{H}_{R}}\left\vert \psi \right\rangle \left\langle \psi
\right\vert $ in terms of a purification $\left\vert \psi \right\rangle \in 
\mathcal{H}\otimes \mathcal{H}_{R}$ to a reference system $\mathcal{H}_{R}$.
The purification $\left\vert \psi \right\rangle $ encodes all of the
information in $\rho $. Entanglement fidelity is a measure of how well the
channel $\Lambda $ preserves the entanglement of the state $\mathcal{H}$
with its reference system $\mathcal{H}_{R}$. The entanglement fidelity is
defined as follows \cite{schumacher96},%
\begin{equation}
\mathcal{F}\left( \rho \text{, }\Lambda \right) \overset{\text{def}}{=}%
\left\langle \psi |\left( \Lambda \otimes I_{\mathcal{H}_{R}}\right) \left(
\left\vert \psi \right\rangle \left\langle \psi \right\vert \right) |\psi
\right\rangle \text{,}
\end{equation}%
where $\left\vert \psi \right\rangle $ is any purification of $\rho $, $I_{%
\mathcal{H}_{R}}$ is the identity map on $\mathcal{M}\left( \mathcal{H}%
_{R}\right) $ and $\Lambda \otimes I_{\mathcal{H}_{R}}$ is the evolution
operator extended to the space $\mathcal{H}\otimes \mathcal{H}_{R}$, space
on which $\rho $ has been purified. If the quantum operation $\Lambda $ is
written in terms of its Kraus error operators $\left\{ A_{k}\right\} $ as, $%
\Lambda \left( \rho \right) =\sum_{k}A_{k}\rho A_{k}^{\dagger }$, then it
can be shown that \cite{nielsen96}, 
\begin{equation}
\mathcal{F}\left( \rho \text{, }\Lambda \right) =\sum_{k}\text{tr}\left(
A_{k}\rho \right) \text{tr}\left( A_{k}^{\dagger }\rho \right)
=\sum_{k}\left\vert \text{tr}\left( \rho A_{k}\right) \right\vert ^{2}\text{.%
}
\end{equation}%
This expression for the entanglement fidelity is very useful for explicit
calculations. Finally, assuming that%
\begin{equation}
\Lambda :\mathcal{M}\left( \mathcal{H}\right) \ni \rho \longmapsto \Lambda
\left( \rho \right) =\sum_{k}A_{k}\rho A_{k}^{\dagger }\in \mathcal{M}\left( 
\mathcal{H}\right) \text{, dim}_{%
\mathbb{C}
}\mathcal{H=}N  \label{pla1}
\end{equation}%
and choosing a purification described by a maximally entangled unit vector $%
\left\vert \psi \right\rangle \in \mathcal{H}\otimes \mathcal{H}$ for the
mixed state $\rho =\frac{1}{\text{dim}_{%
\mathbb{C}
}\mathcal{H}}I_{\mathcal{H}}$ , we obtain%
\begin{equation}
\mathcal{F}\left( \frac{1}{N}I_{\mathcal{H}}\text{, }\Lambda \right) =\frac{1%
}{N^{2}}\sum_{k}\left\vert \text{tr}A_{k}\right\vert ^{2}\text{.}
\label{nfi}
\end{equation}%
The expression\ in (\ref{nfi}) represents the entanglement fidelity when no
error correction is performed on the noisy channel $\Lambda $ in (\ref{pla1}%
).

Here we want to describe the action of $\mathcal{R\circ }\Lambda ^{(5)}$ in (%
\ref{r5}) restricted to the code subspace $\mathcal{C}$. Note that the
recovery operators can be expressed as,%
\begin{equation}
R_{l+1}=R_{1}\frac{A_{l}^{\prime }}{\sqrt{\tilde{p}_{l}}}=\left( \left\vert
0_{L}\right\rangle \left\langle 0_{L}\right\vert +\left\vert
1_{L}\right\rangle \left\langle 1_{L}\right\vert \right) \frac{A_{l}^{\prime
}}{\sqrt{\tilde{p}_{l}}}\text{,}
\end{equation}%
with $l\in \left\{ 0\text{,..., }15\right\} $. Recalling that $A_{l}^{\prime
}=A_{l}^{\prime \dagger }$, it turns out that,%
\begin{equation}
\left\langle i_{L}|R_{l+1}A_{k}^{\prime }|j_{L}\right\rangle =\frac{1}{\sqrt{%
\tilde{p}_{l}}}\left\langle i_{L}|0_{L}\right\rangle \left\langle
0_{L}|A_{l}^{\prime \dagger }A_{k}^{\prime }|j_{L}\right\rangle +\frac{1}{%
\sqrt{\tilde{p}_{l}}}\left\langle i_{L}|1_{L}\right\rangle \left\langle
1_{L}|A_{l}^{\prime \dagger }A_{k}^{\prime }|j_{L}\right\rangle \text{.}
\end{equation}%
We now need to compute the $2\times 2$ matrix representation $\left[
R_{l}A_{k}^{\prime }\right] _{|\mathcal{C}}$ of each $R_{l}A_{k}^{\prime }$
with $l=0$,..., $15$ and $k=0$,..., $2^{10}-1$ where,%
\begin{equation}
\left[ R_{l+1}A_{k}^{\prime }\right] _{|\mathcal{C}}\overset{\text{def}}{=}%
\left( 
\begin{array}{cc}
\left\langle 0_{L}|R_{l+1}A_{k}^{\prime }|0_{L}\right\rangle & \left\langle
0_{L}|R_{l+1}A_{k}^{\prime }|1_{L}\right\rangle \\ 
\left\langle 1_{L}|R_{l+1}A_{k}^{\prime }|0_{L}\right\rangle & \left\langle
1_{L}|R_{l+1}A_{k}^{\prime }|1_{L}\right\rangle%
\end{array}%
\right) \text{.}
\end{equation}%
For $l$, $k=0$,.., $15$, we note that $\left[ R_{l+1}A_{k}^{\prime }\right]
_{|\mathcal{C}}$ becomes,%
\begin{equation}
\left[ R_{l+1}A_{k}^{\prime }\right] _{|\mathcal{C}}=\left( 
\begin{array}{cc}
\left\langle 0_{L}|A_{l}^{\prime \dagger }A_{k}^{\prime }|0_{L}\right\rangle
& 0 \\ 
0 & \left\langle 1_{L}|A_{l}^{\prime \dagger }A_{k}^{\prime
}|1_{L}\right\rangle%
\end{array}%
\right) =\sqrt{\tilde{p}_{l}}\delta _{lk}\left( 
\begin{array}{cc}
1 & 0 \\ 
0 & 1%
\end{array}%
\right) \text{,}
\end{equation}%
while for any pair $\left( l\text{, }k\right) $ with $l$, $=0$,..., $15$ and 
$k>15$, it follows that,%
\begin{equation}
\left\langle 0_{L}|R_{l+1}A_{k}^{\prime }|0_{L}\right\rangle +\left\langle
1_{L}|R_{l+1}A_{k}^{\prime }|1_{L}\right\rangle =0\text{.}
\end{equation}%
We conclude that the only matrices $\left[ R_{l}A_{k}^{\prime }\right] _{|%
\mathcal{C}}$ with non-vanishing trace are given by,%
\begin{equation}
\left[ R_{s}A_{s-1}^{\prime }\right] _{|\mathcal{C}}=\sqrt{\tilde{p}_{s-1}}%
\left( 
\begin{array}{cc}
1 & 0 \\ 
0 & 1%
\end{array}%
\right) \text{, }
\end{equation}%
with $s=1$,.., $16$. Therefore, the entanglement fidelity $\mathcal{F}^{%
\left[ \left[ 5,1,3\right] \right] }\left( \mu \text{, }p\right) $ defined
as,%
\begin{equation}
\mathcal{F}^{\left[ \left[ 5,1,3\right] \right] }\left( \mu \text{, }%
p\right) \overset{\text{def}}{=}\mathcal{F}^{\left[ \left[ 5,1,3\right] %
\right] }\left( \frac{1}{2}I_{2\times 2}\text{, }\mathcal{R\circ }\Lambda
^{(5)}\right) =\frac{1}{\left( 2\right) ^{2}}\sum_{k=0}^{2^{10}-1}\sum%
\limits_{l=1}^{16}\left\vert \text{tr}\left( \left[ R_{l}A_{k}^{\prime }%
\right] _{|\mathcal{C}}\right) \right\vert ^{2}\text{,}  \label{fidel}
\end{equation}%
results,%
\begin{equation}
\mathcal{F}^{\left[ \left[ 5,1,3\right] \right] }\left( \mu \text{, }%
p\right) =\sum_{m=0}^{15}\tilde{p}_{m}=p_{00}^{4}p_{0}+3\left[
2p_{00}^{3}p_{10}p_{0}+3p_{00}^{2}p_{01}p_{10}p_{0}\right] \text{.}
\label{sym5}
\end{equation}%
Notice that the expression in (\ref{fidel}) represents the entanglement
fidelity after the error correction scheme provided by the five-qubit code
is performed on the noisy channel $\Lambda ^{(5)}$. The explicit expression
for $\mathcal{F}^{\left[ \left[ 5,1,3\right] \right] }\left( \mu \text{, }%
p\right) $ in (\ref{sym5}) appears in Appendix A.


\begin{figure}
\begin{center}
\includegraphics[width=0.4\textwidth]{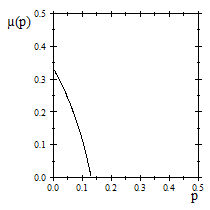}
\end{center}
\vspace{-0.5cm}
\caption{\label{fig1} 
Threshold curve for the
five-qubit code.}
\end{figure}


Note that for arbitrary memory
parameter $\mu $,%
\begin{equation}
\lim_{p\rightarrow 0}\mathcal{F}^{\left[ \left[ 5,1,3\right] \right] }\left(
\mu \text{, }p\right) =1\text{ and, }\lim_{p\rightarrow 1}\mathcal{F}^{\left[
\left[ 5,1,3\right] \right] }\left( \mu \text{, }p\right) =0\text{,}
\end{equation}%
and for $\mu =0$, 
\begin{equation}
\mathcal{F}^{\left[ \left[ 5,1,3\right] \right] }\left( 0\text{, }p\right)
=4p^{5}-15p^{4}+20p^{3}-10p^{2}+1\text{.}
\end{equation}%
We recall that in general the application of a quantum error correcting code
will lower the error probability as long as the probability of error on an
unencoded qubit is less than a certain critical value (threshold
probability). This threshold probability value depends on the code and above
such critical value, the use of a coding scheme only makes the information
corruption worse. Obviously, in order to make effective use of quantum error
correction, a physical implementation of a channel with a sufficiently low
error probability, as well as a code with a sufficiently high threshold is
needed. For instance, the three-qubit repetition code improves the
transmission accuracy when the probability of a bit flip on each qubit sent
through the underlying channel is less than $0.5$. For greater error
probabilities, the error correction process is actually more likely to
corrupt the data than unencoded transmission would be. In our analysis, the
failure probability is represented by \cite{gaitan08},%
\begin{equation}
\mathcal{P}\left( \mu \text{, }p\right) \overset{\text{def}}{=}1-\mathcal{F}%
\left( \mu \text{, }p\right) \text{,}
\end{equation}%
and it gives us an upper bound on the probability with which a generic
encoded state will end up at a wrong state. Therefore, the five-qubit code
is effective only if $\mathcal{P}^{\left[ \left[ 5,1,3\right] \right]
}\left( \mu \text{, }p\right) <p$. The effectiveness parametric region $%
\mathcal{D}^{\left[ \left[ 5,1,3\right] \right] }$ for the five-qubit code
is,%
\begin{equation}
\mathcal{D}^{\left[ \left[ 5,1,3\right] \right] }\overset{\text{def}}{=}%
\left\{ \left( \mu \text{, }p\right) \in \left[ 0\text{, }1\right] \times %
\left[ 0\text{, }1\right] :\mathcal{P}^{\left[ \left[ 5,1,3\right] \right]
}\left( \mu \text{, }p\right) <p\right\} \text{.}
\end{equation}%
For the five-qubit code applied for the correction of correlated
depolarizing errors, it turns out that for increasing values of the memory
parameter $\mu $, the maximum values of the errors probabilities $p$ for
which the correction scheme is effective decrease. More generally, the
threshold curve $\mu _{\text{threshold}}^{\left[ \left[ 5,1,3\right] \right]
}=\mu _{\text{threshold}}^{\left[ \left[ 5,1,3\right] \right] }\left(
p\right) $ defining the parametric region where QEC is effective is plotted
in Figure $1$. Furthermore, we point out that the presence of correlations
in symmetric depolarizing errors does not improve the performance of the
five-qubit code since $\mathcal{F}^{\left[ \left[ 5,1,3\right] \right]
}\left( \mu \text{, }p\right) \leq \mathcal{F}^{\left[ \left[ 5,1,3\right] %
\right] }\left( 0\text{, }p\right) $ for those $\left( \mu \text{, }p\right) 
$-pairs belonging to the parametric region $\mathcal{D}^{^{\left[ \left[
5,1,3\right] \right] }}$. Finally, the plots of $\mathcal{F}^{\left[ \left[
5,1,3\right] \right] }\left( \mu \text{, }p\right) $ vs. $\mu $ for $%
p=4.33\times 10^{-2}$ , $p=4\times 10^{-2}$ and $p=3.67\times 10^{-2}$ are
presented in Figure $2$. 

\begin{figure}
\begin{center}
\includegraphics[width=0.4\textwidth]{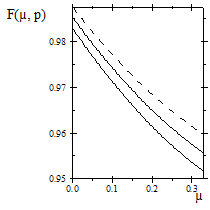}
\end{center}
\vspace{-0.5cm}
\caption{\label{fig2} 
$\mathcal{F}^{\left[ \left[ 5,1,3\right] \right] }\left( \protect\mu \text{, }%
p\right) $ vs. $\protect\mu $ with $0\leq \protect\mu \leq 0.33$ ( for $%
\protect\mu >0.33$, the error correction scheme is not effective anymore)
for $p=4.33\times 10^{-2}$ (thick solid line), $p=4\times 10^{-2}$ (thin
solid line) and $p=3.67\times 10^{-2}$ (dashed line).}
\end{figure}

\section{The Seven-Qubit Code: Symmetric Error Probabilities and Correlations%
}

In this Section, we consider a depolarizing noisy quantum memory channel
with symmetric error probabilities and QEC is performed via the $\left[ %
\left[ 7\text{, }1\text{, }3\right] \right] $-CSS stabilizer code. The
performance of quantum error correcting codes is quantified by means of the
entanglement fidelity $\mathcal{F}^{\left[ \left[ 7,1,3\right] \right]
}\left( \mu \text{, }p\right) $ as function of the error probability $p$ and
degree of memory $\mu $.

\emph{Error Model}. Consider seven qubits and correlated errors in a
depolarizing quantum channel $\Lambda ^{(7)}(\rho )$, 
\begin{equation}
\Lambda ^{(7)}(\rho )=\sum_{i_{1}\text{, }i_{2}\text{, }i_{3}\text{, }i_{4}%
\text{, }i_{5}\text{, }i_{6}\text{, }i_{7}\text{ }%
=0}^{3}p_{i_{7}|i_{6}}p_{i_{6}|i_{5}}p_{i_{5}|i_{4}}p_{i_{4}|i_{3}}p_{i_{3}|i_{2}}p_{i_{2}|i_{1}}p_{i_{1}}%
\left[ A_{i_{7}}A_{i_{6}}A_{i_{5}}A_{i_{4}}A_{i_{3}}A_{i_{2}}A_{i_{1}}\rho
A_{i_{1}}^{\dag }A_{i_{2}}^{\dag }A_{i_{3}}^{\dag }A_{i_{4}}^{\dagger
}A_{i_{5}}^{\dagger }A_{i_{6}}^{\dagger }A_{i_{7}}^{\dagger }\right] \text{,}
\end{equation}%
where $A_{0}\equiv I$, $A_{1}\equiv X$, $A_{2}\equiv Y$, $A_{3}\equiv Z$ are
the Pauli operators and the coefficients $p_{i_{l}i_{m}}$ with $l$, $m$ $\in
\left\{ 0\text{, }1\text{,..., }7\right\} $ satisfying the normalization
condition,%
\begin{equation}
\sum_{i_{1}\text{, }i_{2}\text{, }i_{3}\text{, }i_{4}\text{, }i_{5}\text{, }%
i_{6}\text{, }i_{7}\text{ }%
=0}^{3}p_{i_{7}|i_{6}}p_{i_{6}|i_{5}}p_{i_{5}|i_{4}}p_{i_{4}|i_{3}}p_{i_{3}|i_{2}}p_{i_{2}|i_{1}}p_{i_{1}}=1%
\text{.}
\end{equation}%
For the depolarizing channel $\Lambda ^{(7)}(\rho )$, coefficients $%
p_{i_{l}|i_{m}}$ are explicitly defined in (\ref{conditional}).

\emph{Error Operators}. In an explicit way, the depolarizing channel $%
\Lambda ^{(7)}(\rho )$ can be written as,%
\begin{equation}
\Lambda ^{(7)}(\rho )=\sum_{k=0}^{2^{14}-1}A_{k}^{\prime }\rho A_{k}^{\prime
\dagger }\text{,}  \label{jfk}
\end{equation}%
where $A_{k}^{\prime }$ are the enlarged error operators acting on the seven
qubit quantum states. The cardinality of the error operators defining $%
\Lambda ^{(7)}(\rho )$ is $2^{14}$ and is obtained by noticing that,%
\begin{equation}
\sum_{m=0}^{7}3^{m}\binom{7}{m}=2^{14}\text{,}
\end{equation}%
where $3^{m}\binom{7}{m}$ is the cardinality of weight-$m$ error operators $%
A_{k}^{\prime }$ in (\ref{jfk}).

\emph{Encoding}. The Calderbank-Shor-Steane (CSS) codes are constructed from
two classical binary codes $\mathcal{C}$ and $\mathcal{C}^{\prime }$ that
have the following properties \cite{calderbank-shor96, steane96}: 1) $%
\mathcal{C}$ and $\mathcal{C}^{\prime }$ are $\left[ n\text{, }k\text{, }d%
\right] $ and $\left[ n\text{, }k^{\prime }\text{, }d^{\prime }\right] $
codes, respectively; 2) $\mathcal{C}^{\prime }\subset \mathcal{C}$; 3) $%
\mathcal{C}$ and $\mathcal{C}_{\perp }^{\prime }$ (the dual code of $%
\mathcal{C}^{\prime }$) are both $t$-error correcting codes. For instance,
in case of the seven-qubit code, the two classical codes are the $\left[ 7%
\text{, }4\text{, }3\right] $ binary Hamming code $\left( \mathcal{C}\right) 
$ and the $\left[ 7\text{, }3\text{, }4\right] $ binary simplex code $\left( 
\mathcal{C}^{\prime }\right) $. The dual code $\mathcal{C}_{\perp }^{\prime
} $ is the $\left[ 7\text{, }4\text{, }3\right] $ binary Hamming code. Thus $%
\mathcal{C}$ and $\mathcal{C}_{\perp }^{\prime }$ are both $1$-error
correcting codes. In this case, $n=7$, $k=4$, $k^{\prime }=3$, $k-k^{\prime
}=1$ so that $1$ qubit is mapped into $7$ qubits. The seven-qubit code is
the simplest example of a CSS code. The five-qubit code introduced in the
previous Section is the shortest possible quantum code to correct one error
and is therefore of immense interest. Although the seven-qubit code is
ostensibly more complicated that the five-qubit code, it is actually more
useful in certain situations by virtue of being a CSS\ code. The CSS codes
are a particularly interesting class of codes for two reasons. First, they
are built using classical codes which have been more heavily studied than
quantum codes, so it is fairly easy to construct useful quantum codes simply
by looking at lists of classical codes. Second, because of the form of
generators, the CSS\ codes are precisely those for which a CNOT applied
between every pair of corresponding qubits in two blocks performs a valid
fault-tolerant operation. This makes them particularly good candidates in
fault-tolerant computation.

The $\left[ \left[ 7\text{, }1\text{, }3\right] \right] $-CSS code encodes $%
k=1$ qubit in $n=7$ qubits. The cardinality of its stabilizer group $%
\mathcal{S}$ is $\left\vert \mathcal{S}\right\vert =2^{n-k}=64$ and the set $%
\mathcal{B}_{\mathcal{S}}^{\left[ \left[ 7,1,3\right] \right] }$ of $n-k=6$
group generators is given by \cite{nielsen00},%
\begin{equation}
\mathcal{B}_{\mathcal{S}}^{\left[ \left[ 7,1,3\right] \right] }\overset{%
\text{def}}{=}\left\{ X^{4}X^{5}X^{6}X^{7}\text{, }X^{2}X^{3}X^{6}X^{7}\text{%
, }X^{1}X^{3}X^{5}X^{7}\text{, }Z^{4}Z^{5}Z^{6}Z^{7}\text{, }%
Z^{2}Z^{3}Z^{6}Z^{7}\text{, }Z^{1}Z^{3}Z^{5}Z^{7}\right\} \text{.}
\end{equation}%
The distance of the code is $d=3$ and therefore the weight of the smallest
error $A_{l}^{\prime \dagger }A_{k}^{\prime }$ \ that cannot be detected by
the code is $3$. Finally, we recall that it is a non-degenerate code since
the smallest weight for elements of $\mathcal{S}$ (other than identity) is $%
4 $ and therefore it is greater than the distance $d=3$. The encoding for
the $\left[ \left[ 7\text{, }1\text{, }3\right] \right] $ code is given by 
\cite{nielsen00},%
\begin{equation}
\left\vert 0\right\rangle \rightarrow \left\vert 0_{L}\right\rangle =\frac{1%
}{\left( \sqrt{2}\right) ^{3}}\left[ 
\begin{array}{c}
\left\vert 0000000\right\rangle +\left\vert 0110011\right\rangle +\left\vert
1010101\right\rangle +\left\vert 1100110\right\rangle + \\ 
\\ 
+\left\vert 0001111\right\rangle +\left\vert 0111100\right\rangle
+\left\vert 1011010\right\rangle +\left\vert 1101001\right\rangle%
\end{array}%
\right] \text{,}  \label{code7}
\end{equation}%
and,%
\begin{equation}
\left\vert 1\right\rangle \rightarrow \left\vert 1_{L}\right\rangle =\frac{1%
}{\left( \sqrt{2}\right) ^{3}}\left[ 
\begin{array}{c}
\left\vert 1111111\right\rangle +\left\vert 1001100\right\rangle +\left\vert
0101010\right\rangle +\left\vert 0011001\right\rangle + \\ 
\\ 
+\left\vert 1110000\right\rangle +\left\vert 1000011\right\rangle
+\left\vert 0100101\right\rangle +\left\vert 0010110\right\rangle%
\end{array}%
\right] \text{.}
\end{equation}%
\emph{Recovery Operators}. Recall that errors with non-vanishing error
syndrome are detectable and that a set of invertible error operators $%
\mathcal{A}_{\text{correctable }}$ is correctable if the set given by $%
\mathcal{A}_{\text{correctable }}^{\dagger }\mathcal{A}_{\text{correctable }%
} $is detectable \cite{knill02}. It is straightforward, though tedious, to
check that,%
\begin{equation}
S\left( A_{l}^{\prime \dagger }A_{k}^{\prime }\right) \neq 0\text{, with }l%
\text{, }k\in \left\{ 0\text{, }1\text{,..., }63\right\} \text{,}
\end{equation}%
where $S\left( A_{k}^{\prime }\right) $ is the error syndrome of the error
operator $A_{k}^{\prime }$ (see Appendix B for their explicit expressions)
defined as \cite{gaitan08},%
\begin{equation}
S\left( A_{k}^{\prime }\right) \overset{\text{def}}{=}H^{\left[ \left[ 7,1,3%
\right] \right] }v_{A_{k}^{\prime }}\text{.}
\end{equation}%
The quantity $H^{\left[ \left[ 7,1,3\right] \right] }$ is the check matrix
for the seven-qubit code \cite{nielsen00},%
\begin{equation}
H^{\left[ \left[ 7,1,3\right] \right] }\overset{\text{def}}{=}\left( 
\begin{array}{ccccccccccccccc}
1 & 1 & 1 & 1 & 0 & 0 & 0 & | & 0 & 0 & 0 & 0 & 0 & 0 & 0 \\ 
1 & 1 & 0 & 0 & 1 & 1 & 0 & | & 0 & 0 & 0 & 0 & 0 & 0 & 0 \\ 
1 & 0 & 1 & 0 & 1 & 0 & 1 & | & 0 & 0 & 0 & 0 & 0 & 0 & 0 \\ 
0 & 0 & 0 & 0 & 0 & 0 & 0 & | & 1 & 1 & 1 & 1 & 0 & 0 & 0 \\ 
0 & 0 & 0 & 0 & 0 & 0 & 0 & | & 1 & 1 & 0 & 0 & 1 & 1 & 0 \\ 
0 & 0 & 0 & 0 & 0 & 0 & 0 & | & 1 & 0 & 1 & 0 & 1 & 0 & 1%
\end{array}%
\right) \text{,}
\end{equation}%
and $v_{A_{k}^{\prime }}$ is the vector in the $14$-dimensional binary
vector space $F_{2}^{14}$ corresponding to the error operator $A_{k}^{\prime
}$. The $\left[ \left[ 7\text{, }1\text{, }3\right] \right] $-code has
distance $3$ and therefore all errors $A^{\prime }\equiv A_{l}^{\prime
\dagger }A_{k}^{\prime }$ with $l$, $k\in \left\{ 0,\text{...}%
,2^{14}-1\right\} $ of weight less than $3$ satisfy the relation,%
\begin{equation}
\left\langle i_{L}|A^{\prime }|j_{L}\right\rangle =\alpha _{A^{\prime
}}\delta _{ij}\text{,}  \label{de}
\end{equation}%
and at least one error of weight $3$ exists that violates it. It is
straightforward, though tedious, to check that all $1$- and $2$-qubit error
operators satisfy this equation (therefore, they are detectable). Instead,
there are $3$-qubit errors that do not satisfy (\ref{de}). For instance, the
error operator $X^{1}X^{2}X^{3}$ is such that $\left\langle
0_{L}|X^{1}X^{2}X^{3}|1_{L}\right\rangle =1\neq 0$. The $\left[ \left[ 7%
\text{, }1\text{, }3\right] \right] $-code corrects arbitrary $1$-qubit
errors, not arbitrary $2$-qubit errors. In Appendix B, we introduce the Set-$%
1$ of correctable errors and explicitly show that they are detectable. It
turns out that the set of correctable error operators is given by,%
\begin{equation}
\mathcal{A}_{\text{correctable}}=\left\{ A_{0}^{\prime }\text{, }%
A_{1}^{\prime }\text{,..., }A_{21}^{\prime }\text{, }A_{22}^{\prime }\text{,
..., }A_{63}^{\prime }\right\} \subseteq \mathcal{A}\text{,}  \label{set1}
\end{equation}%
where the cardinality of $\mathcal{A}$ equals $2^{14}$. All weight-$0$,
weight-$1$ and the $42$ weight-$2$ above-mentioned error operators (see
Appendix B) satisfy the error correction conditions,%
\begin{equation}
\left\langle i_{L}|A_{l}^{\prime \dagger }A_{m}^{\prime }|j_{L}\right\rangle
=\alpha _{lm}^{\prime }\delta _{ij}\text{,}
\end{equation}%
for $l$, $m\in \left\{ 0\text{, }1\text{,..., }63\right\} $ and $i$, $j\in
\left\{ 0\text{, }1\right\} $ with $\left\langle i_{L}|j_{L}\right\rangle
=\delta _{ij}$.

In general, QEC\ protocols are symmetric with respect to the phase and bit
bases and so enable the detection and correction of an equal number of phase
and bit errors. In the CSS\ construction a pair of codes are used, one for
correcting the bit flip errors and the other for correcting the phase flip
errors. These codes can be chosen in such a way that the code correcting the
phase flip errors has a larger distance than that of the code correcting the
bit flip errors. Therefore, the resulting asymmetric quantum code has
different error correcting capability for handling different type of errors.
For instance, we emphasize that for the seven-qubit code there is some
freedom in the selection of the set of correctable errors, even after the
stabilizer generators have been specified \cite{david08}. The seven-qubit
code may be designed to prioritize a certain error over the others (say $Z$
errors over $X$ and $X$ errors over $Y$). For instance, an implementation
which has no possibility at all of a $Y$ error could use a code where the
set of correctable errors was chosen to exclude corrections for $Y$.
Optimizing the seven-qubit code to completely remove the ability to correct
one error could lead to qualitatively different behavior, possibly even
including better threshold values \cite{david08}. Instead, the five-qubit
code (which is not a CSS code) corrects a unique symmetric set of errors. In
what follows, first we will compute $\mathcal{F}_{\text{Set-}1}^{\left[ %
\left[ 7,1,3\right] \right] }\left( \mu \text{, }p\right) $ assuming to
correct the set of errors in (\ref{set1}); second, we will compute $\mathcal{%
F}_{\text{Set-}2}^{\left[ \left[ 7,1,3\right] \right] }\left( \mu \text{, }%
p\right) $ assuming to correct the set of errors where we prioritize $Z$
errors over $X$ and $X$ errors over $Y$.

\subsection{Computation of $\mathcal{F}_{\text{Set-1}}^{\left[ \left[ 7\text{%
, }1\text{, }3\right] \right] }\left( \protect\mu \text{, }p\right) $}

The two \ $64$-dimensional orthogonal subspaces $\mathcal{V}^{0_{L}}$ and $%
\mathcal{V}^{1_{L}}$ of $\mathcal{H}_{2}^{7}$ generated by the action of $%
\mathcal{A}_{\text{correctable}}$ on $\left\vert 0_{L}\right\rangle $ and $%
\left\vert 1_{L}\right\rangle $ are given by,%
\begin{equation}
\mathcal{V}^{0_{L}}=Span\left\{ \left\vert v_{l+1}^{0_{L}}\right\rangle =%
\frac{1}{\sqrt{\tilde{p}_{l}^{\prime }}}A_{l}^{\prime }\left\vert
0_{L}\right\rangle \right\} \text{,}
\end{equation}%
with $l\in \left\{ 0\text{,..., }63\right\} $ and,%
\begin{equation}
\mathcal{V}^{1_{L}}=Span\left\{ \left\vert v_{l+1}^{1_{L}}\right\rangle =%
\frac{1}{\sqrt{\tilde{p}_{l}^{\prime }}}A_{l}^{\prime }\left\vert
1_{L}\right\rangle \right\} \text{,}
\end{equation}%
respectively. Notice that $\left\langle v_{l}^{i_{L}}|v_{l^{\prime
}}^{j_{L}}\right\rangle =\delta _{ll^{\prime }}\delta _{ij}$ with $l$, $%
l^{\prime }\in \left\{ 0\text{,..., }63\right\} $ and $i$, $j\in \left\{ 0%
\text{, }1\right\} $. Therefore, it follows that $\mathcal{V}^{0_{L}}\oplus 
\mathcal{V}^{1_{L}}=\mathcal{H}_{2}^{7}$. The recovery superoperator $%
\mathcal{R}\leftrightarrow \left\{ R_{l}\right\} $ with $l=1$,..., $64$ is
defined as \cite{knill97},%
\begin{equation}
R_{l}\overset{\text{def}}{=}V_{l}\sum_{i=0}^{1}\left\vert
v_{l}^{i_{L}}\right\rangle \left\langle v_{l}^{i_{L}}\right\vert \text{,}
\end{equation}%
where the unitary operator $V_{l}$ is such that $V_{l}\left\vert
v_{l}^{i_{L}}\right\rangle =\left\vert i_{L}\right\rangle $ for $i\in
\left\{ 0\text{, }1\right\} $. Notice that,%
\begin{equation}
R_{l}\overset{\text{def}}{=}V_{l}\sum_{i=0}^{1}\left\vert
v_{l}^{i_{L}}\right\rangle \left\langle v_{l}^{i_{L}}\right\vert =\left\vert
0_{L}\right\rangle \left\langle v_{l}^{0_{L}}\right\vert +\left\vert
1_{L}\right\rangle \left\langle v_{l}^{1_{L}}\right\vert \text{.}
\end{equation}%
It turns out that the $64$ recovery operators are given by,%
\begin{equation}
R_{l+1}=R_{1}\frac{A_{l}^{\prime }}{\sqrt{\tilde{p}_{l}^{\prime }}}=\left(
\left\vert 0_{L}\right\rangle \left\langle 0_{L}\right\vert +\left\vert
1_{L}\right\rangle \left\langle 1_{L}\right\vert \right) \frac{A_{l}^{\prime
}}{\sqrt{\tilde{p}_{l}^{\prime }}}\text{,}
\end{equation}%
with $l\in \left\{ 0\text{,..., }63\right\} $. Notice that $\mathcal{R}%
\leftrightarrow \left\{ R_{l}\right\} $ is a trace preserving quantum
operation, $\sum_{l=1}^{64}R_{l}^{\dagger }R_{l}=I_{128\times 128}$ because $%
\left\{ \left\vert v_{l}^{i_{L}}\right\rangle \right\} $ with $l=1$,..., $64$
and $i_{L}\in \left\{ 0\text{, }1\right\} $ is an orthonormal basis for $%
\mathcal{H}_{2}^{7}$. Finally, the action of this recovery operation $%
\mathcal{R}$ on the map $\Lambda ^{\left( 7\right) }\left( \rho \right) $ in
(\ref{nota}) leads to,%
\begin{equation}
\Lambda _{\text{recover}}^{\left( 7\right) }\left( \rho \right) \equiv
\left( \mathcal{R\circ }\Lambda ^{(7)}\right) \left( \rho \right) \overset{%
\text{def}}{=}\sum_{k=0}^{2^{14}-1}\sum\limits_{l=1}^{64}\left(
R_{l}A_{k}^{\prime }\right) \rho \left( R_{l}A_{k}^{\prime }\right)
^{\dagger }\text{.}
\end{equation}

\emph{Entanglement Fidelity}. We want to describe the action of $\mathcal{%
R\circ }\Lambda ^{(7)}$ restricted to the code subspace $\mathcal{C}$.
Recalling that $A_{l}^{\prime }=A_{l}^{\prime \dagger }$, it turns out that,%
\begin{equation}
\left\langle i_{L}|R_{l+1}A_{k}^{\prime }|j_{L}\right\rangle =\frac{1}{\sqrt{%
\tilde{p}_{l}^{\prime }}}\left\langle i_{L}|0_{L}\right\rangle \left\langle
0_{L}|A_{l}^{\prime \dagger }A_{k}^{\prime }|j_{L}\right\rangle +\frac{1}{%
\sqrt{\tilde{p}_{l}^{\prime }}}\left\langle i_{L}|1_{L}\right\rangle
\left\langle 1_{L}|A_{l}^{\prime \dagger }A_{k}^{\prime }|j_{L}\right\rangle 
\text{.}
\end{equation}%
We now need to compute the $2\times 2$ matrix representation $\left[
R_{l}A_{k}^{\prime }\right] _{|\mathcal{C}}$ of each $R_{l}A_{k}^{\prime }$
with $l=0$,..., $63$ and $k=0$,..., $2^{14}-1$ where,%
\begin{equation}
\left[ R_{l+1}A_{k}^{\prime }\right] _{|\mathcal{C}}\overset{\text{def}}{=}%
\left( 
\begin{array}{cc}
\left\langle 0_{L}|R_{l+1}A_{k}^{\prime }|0_{L}\right\rangle  & \left\langle
0_{L}|R_{l+1}A_{k}^{\prime }|1_{L}\right\rangle  \\ 
\left\langle 1_{L}|R_{l+1}A_{k}^{\prime }|0_{L}\right\rangle  & \left\langle
1_{L}|R_{l+1}A_{k}^{\prime }|1_{L}\right\rangle 
\end{array}%
\right) \text{.}
\end{equation}%
For $l$, $k=0$,..., $63$, we note that $\left[ R_{l+1}A_{k}^{\prime }\right]
_{|\mathcal{C}}$ becomes,%
\begin{equation}
\left[ R_{l+1}A_{k}^{\prime }\right] _{|\mathcal{C}}=\left( 
\begin{array}{cc}
\left\langle 0_{L}|A_{l}^{\prime \dagger }A_{k}^{\prime }|0_{L}\right\rangle 
& 0 \\ 
0 & \left\langle 1_{L}|A_{l}^{\prime \dagger }A_{k}^{\prime
}|1_{L}\right\rangle 
\end{array}%
\right) =\sqrt{\tilde{p}_{l}^{\prime }}\delta _{lk}\left( 
\begin{array}{cc}
1 & 0 \\ 
0 & 1%
\end{array}%
\right) \text{,}
\end{equation}%
while for any pair $\left( l\text{, }k\right) $ with $l$ $=0$,..., $63$ and $%
k>63$, it follows that,%
\begin{equation}
\left\langle 0_{L}|R_{l+1}A_{k}^{\prime }|0_{L}\right\rangle +\left\langle
1_{L}|R_{l+1}A_{k}^{\prime }|1_{L}\right\rangle =0\text{.}
\end{equation}%
We conclude that the only matrices $\left[ R_{l}A_{k}^{\prime }\right] _{|%
\mathcal{C}}$ with non-vanishing trace are given by $\left[
R_{l+1}A_{l}^{\prime }\right] _{|\mathcal{C}}$ with $l$ $=0$,.., $63$ where,%
\begin{equation}
\left[ R_{l+1}A_{l}^{\prime }\right] _{|\mathcal{C}}=\sqrt{\tilde{p}%
_{l}^{\prime }}\left( 
\begin{array}{cc}
1 & 0 \\ 
0 & 1%
\end{array}%
\right) \text{.}
\end{equation}

\begin{figure}
\begin{center}
\includegraphics[width=0.4\textwidth]{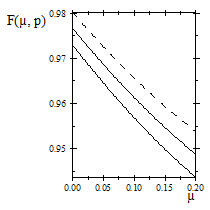}
\end{center}
\vspace{-0.5cm}
\caption{\label{fig3} 
$\mathcal{F}_{\text{Set-}1}^{%
\left[ \left[ 7,1,3\right] \right] }\left( \protect\mu \text{, }p\right) $
vs. $\protect\mu $ for with $0\leq \protect\mu \leq 0.199$ (for $\protect\mu %
>0.199$, the error correction scheme is not effective anymore) $p=4.33\times
10^{-2}$ (thick solid line), $p=4\times 10^{-2}$ (thin solid line) and $%
p=3.67\times 10^{-2}$ (dashed line).}
\end{figure}

Therefore, the entanglement
fidelity $\mathcal{F}_{\text{Set-}1}^{\left[ \left[ 7,1,3\right] \right]
}\left( \mu \text{, }p\right) $ defined as,%
\begin{equation}
\mathcal{F}_{\text{Set-}1}^{\left[ \left[ 7\text{, }1\text{, }3\right] %
\right] }\left( \mu \text{, }p\right) \overset{\text{def}}{=}\mathcal{F}_{%
\text{Set-}1}^{\left[ \left[ 7,1,3\right] \right] }\left( \frac{1}{2}%
I_{2\times 2}\text{, }\mathcal{R\circ }\Lambda ^{(7)}\right) =\frac{1}{%
\left( 2\right) ^{2}}\sum_{k=0}^{2^{14}-1}\sum\limits_{l=1}^{64}\left\vert 
\text{tr}\left( \left[ R_{l}A_{k}^{\prime }\right] _{|\mathcal{C}}\right)
\right\vert ^{2}\text{,}
\end{equation}%
becomes (the explicit expression for $\mathcal{F}_{\text{Set-}1}^{\left[ %
\left[ 7,1,3\right] \right] }\left( \mu \text{, }p\right) $ is given in
Appendix B),%
\begin{equation}
\mathcal{F}_{\text{Set-}1}^{\left[ \left[ 7,1,3\right] \right] }\left( \mu 
\text{, }p\right)
=p_{00}^{6}p_{0}+6p_{00}^{5}p_{10}p_{0}+15p_{00}^{4}p_{01}p_{10}p_{0}+6p_{00}^{4}p_{10}^{2}p_{0}+24p_{00}^{3}p_{01}p_{10}^{2}p_{0}+12p_{00}^{2}p_{01}^{2}p_{10}^{2}p_{0}%
\text{.}  \label{S1}
\end{equation}%
Note that for arbitrary degree of memory $\mu $,%
\begin{equation}
\lim_{p\rightarrow 0}\mathcal{F}_{\text{Set-}1}^{\left[ \left[ 7,1,3\right] %
\right] }\left( \mu \text{, }p\right) =1\text{ and, }\lim_{p\rightarrow 1}%
\mathcal{F}_{\text{Set-}1}^{\left[ \left[ 7,1,3\right] \right] }\left( \mu 
\text{, }p\right) =0\text{,}
\end{equation}%
and for vanishing memory parameter $\mu =0$,%
\begin{equation}
\mathcal{F}_{\text{Set-}1}^{\left[ \left[ 7,1,3\right] \right] }\left( 0%
\text{, }p\right) =\frac{4}{3}p^{7}-\frac{35}{3}p^{6}+\frac{112}{3}p^{5}-%
\frac{175}{3}p^{4}+\frac{140}{3}p^{3}-\frac{49}{3}p^{2}+1\text{.}
\end{equation}%
We emphasize that the presence of correlations in symmetric depolarizing
errors does not improve the performance of the seven-qubit code since $%
\mathcal{F}_{\text{Set-}1}^{\left[ \left[ 7,1,3\right] \right] }\left( \mu 
\text{, }p\right) \leq \mathcal{F}_{\text{Set-}1}^{\left[ \left[ 7,1,3\right]
\right] }\left( 0\text{, }p\right) $ for those $\left( \mu \text{, }p\right) 
$-pairs belonging to the parametric region $\mathcal{D}_{\text{Set-}1}^{^{%
\left[ \left[ 7,1,3\right] \right] }}$ where the correction scheme is
effective,%
\begin{equation}
\mathcal{D}_{\text{Set-}1}^{\left[ \left[ 7,1,3\right] \right] }\overset{%
\text{def}}{=}\left\{ \left( \mu \text{, }p\right) \in \left[ 0\text{, }1%
\right] \times \left[ 0\text{, }1\right] :\mathcal{P}_{\text{Set-}1}^{\left[ %
\left[ 7,1,3\right] \right] }\left( \mu \text{, }p\right) <p\right\} \text{.}
\end{equation}%
Furthermore, it turns out that $\mathcal{F}_{\text{Set-}1}^{\left[ \left[
7,1,3\right] \right] }\left( \mu \text{, }p\right) \leq \mathcal{F}^{\left[ %
\left[ 5,1,3\right] \right] }\left( \mu \text{, }p\right) $ in $\mathcal{D}^{%
\left[ \left[ 5,1,3\right] \right] }\cap \mathcal{D}_{\text{Set-}1}^{\left[ %
\left[ 7,1,3\right] \right] }$ where the area of the parametric region $%
\mathcal{D}_{\text{Set-1}}^{\left[ \left[ 7,1,3\right] \right] }$ is smaller
than that of $\mathcal{D}^{\left[ \left[ 7,1,3\right] \right] }$ (see Figure 
$4$). The plots of $\mathcal{F}_{\text{Set-}1}^{\left[ \left[ 7,1,3\right] %
\right] }\left( \mu \text{, }p\right) $ vs. $\mu $ for $p=4.33\times 10^{-2}$
, $p=4\times 10^{-2}$ and $p=3.67\times 10^{-2}$ appear in Figure $3$. For
the seven-qubit code applied for the correction of correlated depolarizing
errors in Set-$1$, it turns out that for increasing values of the memory
parameter $\mu $, the maximum values of the errors probabilities $p$ for
which the correction scheme is effective decrease. For instance, to $\mu _{%
\text{min}}=0$ corresponds a threshold $p_{\text{threshold}}\cong 7.63\times
10^{-2}$ while to $\mu _{\text{max}}\cong 0.199$ corresponds $p_{\text{%
threshold}}\cong 5.04\times 10^{-4}$.

In the next Subsection, we will study the performance of the seven-qubit
code assuming to correct a new set of correlated error operators. Moreover,
we will compare the performance of the code in such two cases and discuss
the change of the parametric regions where the quantum correction schemes
are effective.

\subsection{Computation of $\mathcal{F}_{\text{Set-2}}^{\left[ \left[ 7\text{%
, }1\text{, }3\right] \right] }\left( \protect\mu \text{, }p\right) $}

Unlike the five-qubit code, the seven-qubit code corrects an asymmetric set
of errors. In what follows, we choose the set of correctable errors to
prioritize $Z$ errors over $X$ errors over $Y$ errors. Said otherwise, we
construct the set of correctable errors by proceeding in increasing order
from single-qubit errors to errors of higher weight. Within each level
(weight) of errors, we include those that incorporate the most $Z$ errors
first. In other words, the sets of weight-$0$ and weight-$1$ correctable
errors are given in (\ref{w0}) and (\ref{w1}), respectively (see Appendix
B). Following the line of reasoning presented in the previous Subsection,
after some algebra it turns out that%
\begin{eqnarray}
\mathcal{F}_{\text{Set-}2}^{\left[ \left[ 7,1,3\right] \right] }\left( \mu 
\text{, }p\right)
&=&p_{00}^{6}p_{0}+6p_{00}^{5}p_{10}p_{0}+15p_{00}^{4}p_{01}p_{10}p_{0}+2p_{00}^{4}p_{10}p_{0}\left( p_{11}+2p_{10}\right) +
\notag \\
&&  \notag \\
&&+4p_{00}^{3}p_{01}p_{10}p_{0}\left( 5p_{10}+p_{11}\right)
+12p_{00}^{2}p_{01}^{2}p_{10}^{2}p_{0}\text{.}  \label{S2}
\end{eqnarray}%
From (\ref{S2}) and (\ref{S1}) (see also Appendix B), we obtain%
\begin{equation}
\mathcal{F}_{\text{Set-}2}^{\left[ \left[ 7,1,3\right] \right] }\left( \mu 
\text{, }p\right) -\mathcal{F}_{\text{Set-}1}^{\left[ \left[ 7\text{, }1%
\text{, }3\right] \right] }\left( \mu \text{, }p\right) =\left(
p_{11}-p_{10}\right) \left(
2p_{00}^{4}p_{10}p_{0}+4p_{00}^{3}p_{01}p_{10}p_{0}\right) \geq 0\text{,}
\end{equation}%
with $\left( p_{11}-p_{10}\right) =\mu \geq 0$. The explicit expression for $%
\mathcal{F}_{\text{Set-}2}^{\left[ \left[ 7,1,3\right] \right] }\left( \mu 
\text{, }p\right) $ appears in Appendix B. In absence of correlations and
considering symmetric error probabilities, the two entanglement fidelities
are the same. Therefore, we conclude that in the presence of memory effects,
it does matter which set of errors we choose to correct, even limiting our
analysis to symmetric error probabilities. We will see that the freedom of
such choice becomes even more important when combining memory effects and
asymmetric error probabilities.

For the seven-qubit code applied for the correction of correlated
depolarizing errors in Set-$2$, it turns out that for increasing values of
the memory parameter $\mu $, the maximum values of the errors probabilities $%
p$ for which the correction scheme is effective decrease. For instance, to $%
\mu _{\text{min}}=0$ corresponds a threshold $p_{\text{threshold}}\cong
7.63\times 10^{-2}$ while to $\mu _{\text{max}}\cong 0.29$ corresponds $p_{%
\text{threshold}}\cong 1.95\times 10^{-3}$.

In conclusion, it follows that in the presence of correlated and symmetric
depolarizing errors, the performances of both the five and the seven-qubit
quantum codes are lowered. Furthermore, the five-qubit code is characterized
by a parametric region (where its correction scheme is effective) that is
larger than the one provided by the seven-qubit code (for both selected sets
of correctable errors). Furthermore, in the parametric region where both
error correction schemes are effective, the five-qubit code outperforms the
seven-qubit code.

In the next Section, we will discover that the situation is slightly
different when considering asymmetries and memory effects in depolarizing
channels. 

\begin{figure}
\begin{center}
\includegraphics[width=0.4\textwidth]{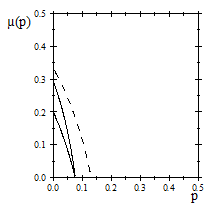}
\end{center}
\vspace{-0.5cm}
\caption{\label{fig4} 
Symmetric Case:
Threshold curves $\protect\mu _{\text{threshold}}^{\left[ \left[ 5,1,3\right]
\right] }\left( p\right) $ (dashed line), $\protect\mu _{\text{threshold,
Set-2}}^{\left[ \left[ 7,1,3\right] \right] }\left( p\right) $ (thin solid
line) and $\protect\mu _{\text{threshold, Set-1}}^{\left[ \left[ 7,1,3\right]
\right] }\left( p\right) $ (thick solid line) vs. $p$.
}
\end{figure}

\section{The Five and Seven-Qubit Codes: Asymmetric Error Probabilities and
Correlations}

In this Section, we study the performance of the $\left[ \left[ 5\text{, }1%
\text{, }3\right] \right] $ and $\left[ \left[ 7\text{, }1\text{, }3\right] %
\right] $ quantum error correcting codes with respect to asymmetric error
probabilities ( $p=p_{X}+p_{Y}+p_{Z}$ with $p_{X}\neq p_{Y}\neq p_{Z}$) and
correlated noise errors in a quantum depolarizing channel.

\subsection{The Five-Qubit Code}

In our following discussion, we will assume that the error probability $p$
may be written as,%
\begin{equation}
p=p_{X}+p_{Y}+p_{Z}\text{,}
\end{equation}%
where,%
\begin{equation}
p_{X}=\alpha _{X}p\text{, }p_{Y}=\alpha _{Y}p\text{, }p_{Z}=\alpha _{Z}p%
\text{,}
\end{equation}%
with $\alpha _{X}+\alpha _{Y}+\alpha _{Z}=1$. Notice that in the symmetric
case, we simply have $\alpha _{X}=\alpha _{Y}=\alpha _{Z}=\frac{1}{3}$.
Following the line of reasoning presented in Section II, it turns out that
the $\mathcal{F}_{\text{asymmetric}}^{\left[ \left[ 5,1,3\right] \right]
}\left( \mu \text{, }p\right) $ becomes,%
\begin{eqnarray}
\mathcal{F}_{\text{asymmetric}}^{\left[ \left[ 5,1,3\right] \right] }\left(
\mu \text{, }p\right) &=&p_{00}^{4}p_{0}+\left[
p_{00}^{3}p_{10}p_{0}+3p_{00}^{2}p_{01}p_{10}p_{0}+p_{00}^{3}p_{01}p_{1}%
\right] +  \notag \\
&&  \notag \\
&&\left[
p_{00}^{3}p_{20}p_{0}+3p_{00}^{2}p_{02}p_{20}p_{0}+p_{00}^{3}p_{02}p_{2}%
\right] +\left[
p_{00}^{3}p_{30}p_{0}+3p_{00}^{2}p_{03}p_{30}p_{0}+p_{00}^{3}p_{03}p_{3}%
\right] \text{,}  \label{asym5}
\end{eqnarray}%
where,%
\begin{eqnarray}
p_{0} &=&1-p\text{, }p_{1}=\alpha _{X}p\text{, }p_{2}=\alpha _{Y}p\text{, }%
p_{3}=\alpha _{Z}p\text{, }p_{00}=\left( 1-\mu \right) \left( 1-p\right)
+\mu \text{,}  \notag \\
&&  \notag \\
p_{01} &=&p_{02}=p_{03}=\left( 1-\mu \right) \left( 1-p\right) \text{, }%
p_{10}=\alpha _{X}p\left( 1-\mu \right) \text{, }  \notag \\
&&  \notag \\
p_{20} &=&\alpha _{Y}p\left( 1-\mu \right) \text{, }p_{30}=\alpha
_{Z}p\left( 1-\mu \right) \text{.}  \label{sup}
\end{eqnarray}%
After some straightforward algebra, $\mathcal{F}_{\text{asymmetric}}^{\left[ %
\left[ 5\text{, }1\text{, }3\right] \right] }\left( \mu \text{, }p\right) $
in (\ref{asym5}) may be written as,%
\begin{equation}
\mathcal{F}_{\text{asymmetric}}^{\left[ \left[ 5,1,3\right] \right] }\left(
\mu \text{, }p\right) =p_{00}^{4}p_{0}+p_{00}^{3}p_{0}\left(
p_{10}+p_{20}+p_{30}\right) +3p_{00}^{2}p_{01}p_{0}\left(
p_{10}+p_{20}+p_{30}\right) +p_{00}^{3}p_{01}\left( p_{1}+p_{2}+p_{3}\right) 
\text{.}  \label{d}
\end{equation}%
Recalling that in the symmetric case $p_{1}=p_{2}=p_{3}=\frac{p}{3}$, $%
p_{10}=p_{20}=p_{30}=\frac{p}{3}\left( 1-\mu \right) $ and substituting (\ref%
{sup}) in (\ref{d}), it follows that%
\begin{equation}
\mathcal{F}_{\text{asymmetric}}^{\left[ \left[ 5,1,3\right] \right] }\left(
\mu \text{, }p\right) =\mathcal{F}_{\text{symmetric}}^{\left[ \left[ 5,1,3%
\right] \right] }\left( \mu \text{, }p\right) \text{.}
\end{equation}%
Therefore, we conclude that the performance of the five-qubit code cannot be
enhanced in the case of asymmetric error probabilities in the depolarizing
channel. This result was somehow expected since the five-qubit code corrects
a unique and symmetric set of error operators.

\subsection{The Seven-Qubit Code}

\begin{figure}
\begin{center}
\includegraphics[width=0.4\textwidth]{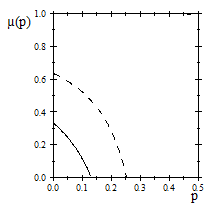}
\end{center}
\vspace{-0.5cm}
\caption{\label{fig5} 
Asymmetric Case: Threshold curves 
$\protect\mu _{\text{threshold, Set-}2}^{\left[ \left[ 7,1,3\right] \right]
}\left( p\right) $ (dashed line) and $\protect\mu _{\text{threshold}}^{\left[
\left[ 5,1,3\right] \right] }\left( p\right) $ (thin solid line) vs. $p$.
}
\end{figure}

Following the line of reasoning
presented in Section III, it turns out that the entanglement fidelity $%
\mathcal{F}_{\text{asymmetric}}^{\left[ \left[ 7,1,3\right] \right] }\left(
\mu \text{, }p\right) $ evaluated assuming to correct the Set-$2$ of error
operators becomes,%
\begin{eqnarray}
\mathcal{F}_{\text{asymmetric}}^{\left[ \left[ 7,1,3\right] \right] }\left(
\mu \text{, }p\right)  &=&p_{00}^{6}p_{0}+2p_{00}^{5}p_{01}\left(
p_{1}+p_{2}+p_{3}\right) +5p_{00}^{4}p_{01}p_{0}\left(
p_{10}+p_{20}+p_{30}\right) +p_{00}^{4}p_{0}\left(
2p_{30}p_{33}+p_{30}^{2}+3p_{10}p_{31}\right) +  \notag \\
&&  \notag \\
&&+p_{00}^{3}p_{01}p_{30}p_{0}\left( 8p_{30}+4p_{33}+12p_{10}\right)
+6p_{00}^{2}p_{01}^{2}p_{30}p_{0}\left( p_{30}+p_{10}\right) \text{,}
\label{xx}
\end{eqnarray}%
where,%
\begin{eqnarray}
p_{0} &=&1-p\text{, }p_{1}=\alpha _{X}p\text{, }p_{2}=\alpha _{Y}p\text{, }%
p_{3}=\alpha _{Z}p\text{, }p_{00}=\left( 1-\mu \right) \left( 1-p\right)
+\mu \text{,}  \notag \\
&&  \notag \\
p_{01} &=&p_{02}=p_{03}=\left( 1-\mu \right) \left( 1-p\right) \text{, }%
p_{10}=\alpha _{X}p\left( 1-\mu \right) \text{, }p_{20}=\alpha _{Y}p\left(
1-\mu \right) \text{, }p_{30}=\alpha _{Z}p\left( 1-\mu \right) \text{,} 
\notag \\
&&  \notag \\
p_{31} &=&p_{30}=\alpha _{Z}p\left( 1-\mu \right) \text{, }p_{33}=\alpha
_{Z}p\left( 1-\mu \right) +\mu \text{.}  \label{x}
\end{eqnarray}%
Notice that for $p_{X}=p_{Y}=p_{Z}=\frac{p}{3}$, $\mathcal{F}_{\text{%
asymmetric}}^{\left[ \left[ 7,1,3\right] \right] }\left( \mu \text{, }%
p\right) $ equals $\mathcal{F}_{\text{symmetric}}^{\left[ \left[ 7,1,3\right]
\right] }\left( \mu \text{, }p\right) $. In absence of correlations, the
entanglement fidelity $\mathcal{F}_{\text{asymmetric}}^{\left[ \left[ 7,1,3%
\right] \right] }$ becomes,%
\begin{equation}
\mathcal{F}_{\text{asymmetric}}^{\left[ \left[ 7,1,3\right] \right] }\left( 0%
\text{, }p\right) =\left( 1-p\right) ^{7}+7p\left( 1-p\right)
^{6}+21p^{2}\left( 1-p\right) ^{5}\left[ \alpha _{Z}^{2}+\alpha _{X}\alpha
_{Z}\right] \text{.}
\end{equation}%
The general expression of $\mathcal{F}_{\text{asymmetric}}^{\left[ \left[
7,1,3\right] \right] }\left( \mu \text{, }p\right) $ is given in Appendix C.
We point out that in the absence of correlations but with asymmetric error
probabilities, the seven-qubit code can outperforms the five-qubit code,%
\begin{equation}
\mathcal{F}_{\text{asymmetric}}^{\left[ \left[ 7,1,3\right] \right] }\left( 0%
\text{, }p\right) \geq \mathcal{F}_{\text{asymmetric}}^{\left[ \left[ 5,1,3%
\right] \right] }\left( 0\text{, }p\right) \equiv \mathcal{F}_{\text{%
symmetric}}^{\left[ \left[ 5,1,3\right] \right] }\left( 0\text{, }p\right) 
\text{.}
\end{equation}%
In Figure $5$, we plot the threshold curves $\mu _{\text{threshold}}^{\left[ %
\left[ 7,1,3\right] \right] }\left( p\right) $ and $\mu _{\text{threshold}}^{%
\left[ \left[ 5,1,3\right] \right] }\left( p\right) $ versus $p$ in the case
case of asymmetric error probabilities. Asymmetries in the error
probabilities enlarge the parametric regions where the seven-qubit code is
effective for error correction. Furthermore, comparing the performances of
such codes on a common region where they are both effective, the seven-qubit
code turns out to outperform the five-qubit code in the presence of
asymmetries and correlations. In Figure $6$, we plot $\mathcal{F}_{\text{%
asymmetric}}^{\left[ \left[ 7,1,3\right] \right] }\left( \mu \text{, }%
p\right) $, $\mathcal{F}_{\text{symmetric}}^{\left[ \left[ 5,1,3\right] %
\right] }\left( \mu \text{, }p\right) =\mathcal{F}_{\text{asymmetric}}^{%
\left[ \left[ 5,1,3\right] \right] }\left( \mu \text{, }p\right) $ and $%
\mathcal{F}_{\text{symmetric}}^{\left[ \left[ 7,1,3\right] \right] }\left(
\mu \text{, }p\right) $ versus the memory parameter $\mu $ for $p=4\times
10^{-2}$ and $\alpha _{Z}=25\alpha $, $\alpha _{X}=5\alpha $ and $\alpha
_{Y}=\alpha $ with $\alpha _{X}+\alpha _{Y}+\alpha _{Z}=1$.

\begin{figure}
\begin{center}
\includegraphics[width=0.4\textwidth]{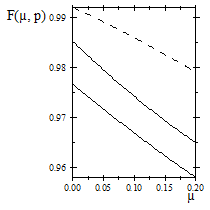}
\end{center}
\vspace{-0.5cm}
\caption{\label{fig6} 
$\mathcal{F}_{\text{asymmetric}}^{\left[ \left[
7,1,3\right] \right] }\left( \protect\mu \text{, }p\right) $ (dashed line), $%
\mathcal{F}_{\text{asymmetric}}^{\left[ \left[ 5,1,3\right] \right] }=%
\mathcal{F}_{\text{symmetric}}^{\left[ \left[ 5,1,3\right] \right] }\left( 
\protect\mu \text{, }p\right) $ (thin solid line) and $\mathcal{F}_{\text{%
symmetric}}^{\left[ \left[ 7,1,3\right] \right] }\left( \protect\mu \text{, }%
p\right) $ (thick solid line) vs. $\protect\mu $ with $0\leq \protect\mu %
\lesssim 0.199$ (where both error correction schemes are effective) for $%
p=4\times 10^{-2}$ with $\protect\alpha _{Z}=25\protect\alpha $, $\protect%
\alpha _{X}=5\protect\alpha $ and $\protect\alpha _{Y}=\protect\alpha $.
}
\end{figure}

\section{Final Remarks}

In this article, we have studied the performance of common quantum
stabilizer codes in the presence of asymmetric and correlated errors.
Specifically, we considered the depolarizing noisy quantum memory channel
and performed quantum error correction via the five and seven-qubit
stabilizer codes. We have shown that memory effects in the error models
combined with asymmetries in the error probabilities can produce relevant
changes in the performances of quantum error correction schemes by
qualitatively affecting the threshold error probability values for which the
codes are effective. In summary, we have uncovered the following findings:

\begin{enumerate}
\item In the presence of correlated and symmetric depolarizing errors, the
performances of both the five and the seven-qubit quantum stabilizer codes
are lowered for fixed values of the degree of memory $\mu $. Furthermore,
such error correction schemes only work for low values of $\mu $.

\item In the presence of correlated and symmetric depolarizing errors, the
five-qubit code is characterized by a parametric region (where its
correction scheme is effective) that is larger than the one provided by the
seven-qubit code. Furthermore, in the parametric region where both error
correction schemes are effective, the five-qubit code outperforms the
seven-qubit code.

\item The asymmetry in the error probabilities does not affect the
performance of the five-qubit code quantified in terms of its entanglement
fidelity. On the contrary, it does affect the performance of the seven-qubit
code which is less effective when considering correlated and symmetric
depolarizing errors. This peculiar effect is rooted in the stabilizer
structure of the CSS seven-qubit code: it is a consequence of the freedom in
selecting the set of correctable error operators even after the stabilizer
generators have been specified.

\item The performance of the seven-qubit code significantly improves when
considering correlated and asymmetric depolarizing errors. Furthermore, in
such case it is also characterized by higher (than the one provided by the
five-qubit code) error probability threshold values. This result confirms
that in order to make effective use of quantum error correction, a physical
implementation of a channel with a sufficiently low error probability, as
well as a code with sufficiently high threshold probability is needed \cite%
{knill05}. \ 
\end{enumerate}

We conclude that in order to optimize the seven-qubit code performance, it
is very important to know the experimental details of the physical
implementation of the quantum memory channel being considered. Furthermore,
in order to make effective use of quantum error correction, a more detailed
analysis of the physical noise models for various qubit implementations is
needed. This requirement, as we have shown, becomes even more pressing when
dealing with noise models where memory effects are combined with asymmetries
in the error probabilities.

\begin{acknowledgments}
C. C. thanks C. Lupo and L. Memarzadeh for useful comments. C. C. is also
grateful to V. Aggarwal and R. Calderbank for their kind hospitality and
useful discussions during his visit at The Program in Applied and
Computational Mathematics at Princeton University. This work was supported
by the European Community's Seventh Framework Program under grant agreement
213681 (CORNER Project; FP7/2007-2013).
\end{acknowledgments}

\appendix

\section{The Five-Qubit Code}

In this Appendix, we briefly discuss few technical details omitted in the
manuscript concerning the application of the five-qubit code to the
depolarizing memory channel with both symmetric and asymmetric error
probabilities.

\emph{Error Operators}. The weight zero and one quantum error operators in $%
\Lambda ^{(5)}(\rho )$ in (\ref{nota}) are given by,%
\begin{eqnarray}
A_{0}^{\prime } &=&\sqrt{\tilde{p}_{0}}I^{1}\otimes I^{2}\otimes
I^{3}\otimes I^{4}\otimes I^{5}\text{, }A_{1}^{\prime }=\sqrt{\tilde{p}_{1}}%
X^{1}\otimes I^{2}\otimes I^{3}\otimes I^{4}\otimes I^{5}\text{, }%
A_{2}^{\prime }=\sqrt{\tilde{p}_{2}}I^{1}\otimes X^{2}\otimes I^{3}\otimes
I^{4}\otimes I^{5}\text{,}  \notag \\
&&  \notag \\
A_{3}^{\prime } &=&\sqrt{\tilde{p}_{3}}I^{1}\otimes I^{2}\otimes
X^{3}\otimes I^{4}\otimes I^{5}\text{, }A_{4}^{\prime }=\sqrt{\tilde{p}_{4}}%
I^{1}\otimes I^{2}\otimes I^{3}\otimes X^{4}\otimes I^{5}\text{, }%
A_{5}^{\prime }=\sqrt{\tilde{p}_{5}}I^{1}\otimes I^{2}\otimes I^{3}\otimes
I^{4}\otimes X^{5}\text{,}  \notag \\
&&  \notag \\
A_{6}^{\prime } &=&\sqrt{\tilde{p}_{6}}Y^{1}\otimes I^{2}\otimes
I^{3}\otimes I^{4}\otimes I^{5}\text{, }A_{7}^{\prime }=\sqrt{\tilde{p}_{7}}%
I^{1}\otimes Y^{2}\otimes I^{3}\otimes I^{4}\otimes I^{5}\text{, }%
A_{8}^{\prime }=\sqrt{\tilde{p}_{8}}I^{1}\otimes I^{2}\otimes Y^{3}\otimes
I^{4}\otimes I^{5}\text{,}  \notag \\
&&  \notag \\
A_{9}^{\prime } &=&\sqrt{\tilde{p}_{9}}I^{1}\otimes I^{2}\otimes
I^{3}\otimes Y^{4}\otimes I^{5}\text{, }A_{10}^{\prime }=\sqrt{\tilde{p}_{10}%
}I^{1}\otimes I^{2}\otimes I^{3}\otimes I^{4}\otimes Y^{5}\text{, }%
A_{11}^{\prime }=\sqrt{\tilde{p}_{11}}Z^{1}\otimes I^{2}\otimes I^{3}\otimes
I^{4}\otimes I^{5}\text{,}  \notag \\
&&  \notag \\
A_{12}^{\prime } &=&\sqrt{\tilde{p}_{12}}I^{1}\otimes Z^{2}\otimes
I^{3}\otimes I^{4}\otimes I^{5}\text{, }A_{13}^{\prime }=\sqrt{\tilde{p}_{13}%
}I^{1}\otimes I^{2}\otimes Z^{3}\otimes I^{4}\otimes I^{5}\text{, }%
A_{14}^{\prime }=\sqrt{\tilde{p}_{14}}I^{1}\otimes I^{2}\otimes I^{3}\otimes
Z^{4}\otimes I^{5}\text{,}  \notag \\
&&  \notag \\
A_{15}^{\prime } &=&\sqrt{\tilde{p}_{15}}I^{1}\otimes I^{2}\otimes
I^{3}\otimes I^{4}\otimes Z^{5}\text{,}
\end{eqnarray}%
where the coefficients $\tilde{p}_{l\text{ }}$ with $l=0$,..., $15$ results,%
\begin{eqnarray}
\tilde{p}_{0} &=&p_{00}^{4}p_{0}\text{, }\tilde{p}_{1}=p_{00}^{3}p_{10}p_{0}%
\text{, }\tilde{p}_{2}=p_{00}^{2}p_{01}p_{10}p_{0}\text{, }\tilde{p}%
_{3}=p_{00}^{2}p_{01}p_{10}p_{0}\text{, }\tilde{p}%
_{4}=p_{00}^{2}p_{01}p_{10}p_{0}\text{, }\tilde{p}_{5}=p_{00}^{3}p_{01}p_{1}%
\text{,}  \notag \\
&&  \notag \\
\tilde{p}_{6} &=&p_{00}^{3}p_{20}p_{0}\text{, }\tilde{p}%
_{7}=p_{00}^{2}p_{02}p_{20}p_{0}\text{, }\tilde{p}%
_{8}=p_{00}^{2}p_{02}p_{20}p_{0}\text{, }\tilde{p}%
_{9}=p_{00}^{2}p_{02}p_{20}p_{0}\text{, }\tilde{p}_{10}=p_{00}^{3}p_{02}p_{2}%
\text{,}  \notag \\
&&  \notag \\
\tilde{p}_{11} &=&p_{00}^{3}p_{30}p_{0}\text{, }\tilde{p}%
_{12}=p_{00}^{2}p_{03}p_{30}p_{0}\text{, }\tilde{p}%
_{13}=p_{00}^{2}p_{03}p_{30}p_{0}\text{, }\tilde{p}%
_{14}=p_{00}^{2}p_{03}p_{30}p_{0}\text{, }\tilde{p}%
_{15}=p_{00}^{3}p_{03}p_{3}\text{,}
\end{eqnarray}%
with,%
\begin{eqnarray}
p_{0} &=&1-p\text{, }p_{1}=p_{2}=p_{3}=\frac{p}{3}\text{, }p_{00}=\left(
1-\mu \right) \left( 1-p\right) +\mu \text{, }  \notag \\
&&  \notag \\
p_{01} &=&p_{02}=p_{03}=\left( 1-\mu \right) \left( 1-p\right) \text{, }%
p_{10}=p_{20}=p_{30}=\frac{p}{3}\left( 1-\mu \right) \text{.}
\end{eqnarray}%
\emph{Detectable Errors}. Recall that an error operator $A_{k}^{\prime }$ is
detectable by the code $\mathcal{C}$ if and only if%
\begin{equation}
P_{\mathcal{C}}A_{k}^{\prime }P_{\mathcal{C}}=\lambda _{A_{k}^{\prime }}P_{%
\mathcal{C}}\text{,}  \label{AAA}
\end{equation}%
for some $\lambda _{A_{k}^{\prime }}$ where $P_{\mathcal{C}}\overset{\text{%
def}}{=}\left\vert 0_{L}\right\rangle \left\langle 0_{L}\right\vert
+\left\vert 1_{L}\right\rangle \left\langle 1_{L}\right\vert $ is the
projector on the code space. On the contrary, a set of error operators $%
\mathcal{A}=\left\{ A_{l}^{\prime }\right\} $ is correctable if and only if%
\begin{equation}
P_{\mathcal{C}}A_{m}^{\prime \dagger }A_{n}^{\prime }P_{\mathcal{C}}=\lambda
_{mn}P_{\mathcal{C}}\text{,}  \label{AA}
\end{equation}%
for any pair of error operators in $\mathcal{A}$ where $\lambda _{mn}$
define a positive semi-definite Hermitian matrix. We emphasize that the
notion of \emph{correctability} depends on all the errors in the set under
consideration and, unlike \emph{detectability}, cannot be applied to
individual errors. For invertible error operators (such as the ones
considered here), there is a simple relationship between detectability and
correctability.\ A set $\mathcal{A}$ is correctable if and only if the
operators in the set $\mathcal{A}^{\dagger }\mathcal{A}\overset{\text{def}}{=%
}\left\{ A_{1}^{\prime \dagger }A_{2}^{\prime }:A_{i}^{\prime }\in \mathcal{A%
}\right\} $ are detectable. It would be awfully tedious to identify either
detectable errors or sets of correctable errors by means of (\ref{AAA}) and (%
\ref{AA}) for the five and seven-qubit codes characterized by the codewords
in (\ref{code5}) and (\ref{code7}), respectively. However, the quantum
stabilizer formalism allows to simplify such task. This is a consequence of
the fact that by means of such formalism it is sufficient to study the
effect of the error operators on the generators of the stabilizer and not on
the codewords themselves. In our work, we have made use of the stabilizer
formalism together with the simple relationship between detectability and
correctability for invertible error operators s in order to identify sets of
correctable and detectable errors.

It is known that errors with non-vanishing error syndrome are detectable. It
is straightforward to check that,%
\begin{equation}
S\left( A_{l}^{\prime \dagger }A_{k}^{\prime }\right) \neq 0\text{, with }l%
\text{, }k\in \left\{ 0\text{, }1\text{,..., }15\right\} \text{,}
\label{this}
\end{equation}%
where $S\left( A_{k}^{\prime }\right) $ is the error syndrome of the error
operator $A_{k}^{\prime }$ defined as,%
\begin{equation}
S\left( A_{k}^{\prime }\right) \overset{\text{def}}{=}H^{\left[ \left[ 5,1,3%
\right] \right] }v_{A_{k}^{\prime }}\text{.}
\end{equation}%
The quantity $H^{\left[ \left[ 5,1,3\right] \right] }$ is the check matrix
for the five-qubit code in (\ref{check1}) and $v_{A_{k}^{\prime }}$ is the
vector in the $10$-dimensional binary vector space $F_{2}^{10}$
corresponding to the error operator $A_{k}^{\prime }$. For instance,
considering $k\in \left\{ 0\text{, }1\text{,..., }15\right\} $, we obtain%
\begin{eqnarray}
v_{I} &=&\left( 00000|00000\right) \text{, }v_{X^{1}}=\left(
10000|00000\right) \text{, }v_{X^{2}}=\left( 01000|00000\right) \text{, }%
v_{X^{3}}=\left( 00100|00000\right) \text{,}  \notag \\
&&  \notag \\
\text{ }v_{X^{4}} &=&\left( 00010|00000\right) \text{, }v_{X^{5}}=\left(
00001|00000\right) \text{, }v_{Z^{1}}=\left( 00000|10000\right) \text{, }%
v_{Z^{2}}=\left( 00000|01000\right) \text{, }  \notag \\
&&  \notag \\
v_{Z^{3}} &=&\left( 00000|00100\right) \text{, }v_{Z^{4}}=\left(
00000|00010\right) \text{, }v_{Z^{5}}=\left( 00000|00001\right) \text{, }%
v_{Y^{1}}=\left( 10000|10000\right) \text{, }  \notag \\
&&  \notag \\
v_{Y^{2}} &=&\left( 01000|01000\right) \text{, }v_{Y^{3}}=\left(
00100|00100\right) \text{, }v_{Y^{4}}=\left( 00010|00010\right) \text{, }%
v_{Y^{5}}=\left( 00001|00001\right) \text{,}
\end{eqnarray}%
and the error syndromes become,%
\begin{eqnarray}
S\left( I\right) &=&0000\text{, }S\left( X^{1}\right) =1000\text{, }S\left(
X^{2}\right) =1100\text{, }S\left( X^{3}\right) =0110\text{, }S\left(
X^{4}\right) =0011\text{, }S\left( X^{5}\right) =0001\text{,}  \notag \\
&&  \notag \\
S\left( Z^{1}\right) &=&0101\text{, }S\left( Z^{2}\right) =0010\text{, }%
S\left( Z^{3}\right) =1001\text{, }S\left( Z^{4}\right) =0100\text{, }%
S\left( Z^{5}\right) =1010\text{,}  \notag \\
&&  \notag \\
S\left( Y^{1}\right) &=&1101\text{, }S\left( Y^{2}\right) =1110\text{, }%
S\left( Y^{3}\right) =1111\text{, }S\left( Y^{4}\right) =0111\text{, }%
S\left( Y^{5}\right) =1011\text{.}  \label{sindrome}
\end{eqnarray}%
For a non-degenerate quantum stabilizer code, linearly independent
correctable errors have unequal error syndromes. This necessary (but not
sufficient) requirement for a set of correctable errors appears fulfilled in
(\ref{sindrome}). Finally, following the above-mentioned line of reasoning,
we can show that (\ref{this}) is fulfilled.

\emph{Recovery Operators}. From (\ref{may1}), it follows that the sixteen
recovery operators are given by,%
\begin{eqnarray}
R_{1} &=&\left\vert 0_{L}\right\rangle \left\langle 0_{L}\right\vert
+\left\vert 1_{L}\right\rangle \left\langle 1_{L}\right\vert \text{, }%
R_{2}=R_{1}X^{1}\text{, }R_{3}=R_{1}X^{2}\text{, }R_{4}=R_{1}X^{3}\text{, }%
R_{5}=R_{1}X^{4}\text{, }R_{6}=R_{1}X^{5}\text{, }  \notag \\
&&  \notag \\
R_{7} &=&R_{1}Y^{1}\text{, }R_{8}=R_{1}Y^{2}\text{, }R_{9}=R_{1}Y^{3}\text{, 
}R_{10}=R_{1}Y^{4}\text{, }R_{11}=R_{1}Y^{5}\text{,}  \notag \\
&&  \notag \\
R_{12} &=&R_{1}Z^{1}\text{, }R_{13}=R_{1}Z^{2}\text{, }R_{14}=R_{1}Z^{3}%
\text{, }R_{15}=R_{1}Z^{4}\text{, }R_{16}=R_{1}Z^{5}\text{.}
\end{eqnarray}

\emph{Entanglement Fidelity}. The explicit expression for the entanglement
fidelity $\mathcal{F}^{\left[ \left[ 5,1,3\right] \right] }\left( \mu \text{%
, }p\right) $ in (\ref{sym5}) is given by,%
\begin{eqnarray}
\mathcal{F}^{\left[ \left[ 5,1,3\right] \right] }\left( \mu \text{, }%
p\right) &=&\mu ^{4}\left( 4p^{5}-7p^{4}+3p^{3}\right) +\mu ^{3}\left(
-16p^{5}+36p^{4}-26p^{3}+6p^{2}\right) +  \notag \\
&&  \notag \\
&&+\mu ^{2}\left( 24p^{5}-66p^{4}+63p^{3}-24p^{2}+3p\right) +\mu \left(
-16p^{5}+52p^{4}-60p^{3}+28p^{2}-4p\right) +  \notag \\
&&  \notag \\
&&+\left( 4p^{5}-15p^{4}+20p^{3}-10p^{2}+1\right) \text{.}  \label{use21}
\end{eqnarray}

\section{On the Seven-Qubit Code}

In this Appendix, we briefly discuss few technical details omitted in the
manuscript concerning the application of the seven-qubit code to the
depolarizing memory channel with both symmetric and asymmetric error
probabilities.

\emph{Error Operators }(Set-$1$). The set of correctable error operators
(Set-$1$) is explicitly defined by $64$ error operators. The only weight-$0$
error operator is given by,%
\begin{equation}
A_{0}^{\prime }=\sqrt{\tilde{p}_{0}^{\prime }}I^{1}\otimes I^{2}\otimes
I^{3}\otimes I^{4}\otimes I^{5}\otimes I^{6}\otimes I^{7}\equiv \sqrt{\tilde{%
p}_{0}^{\prime }}I\text{.}  \label{w0}
\end{equation}%
The $21$ weight-$1$ error (correctable) operators are given by,%
\begin{eqnarray}
A_{1}^{\prime } &=&\sqrt{\tilde{p}_{1}^{\prime }}X^{1}\text{, }A_{2}^{\prime
}=\sqrt{\tilde{p}_{2}^{\prime }}X^{2}\text{, }A_{3}^{\prime }=\sqrt{\tilde{p}%
_{3}^{\prime }}X^{3}\text{, }A_{4}^{\prime }=\sqrt{\tilde{p}_{4}^{\prime }}%
X^{4}\text{, }A_{5}^{\prime }=\sqrt{\tilde{p}_{5}^{\prime }}X^{5}\text{, } 
\notag \\
&&  \notag \\
A_{6}^{\prime } &=&\sqrt{\tilde{p}_{6}^{\prime }}X^{6}\text{, }A_{7}^{\prime
}=\sqrt{\tilde{p}_{7}^{\prime }}X^{7}\text{, }A_{8}^{\prime }=\sqrt{\tilde{p}%
_{8}^{\prime }}Y^{1}\text{, }A_{9}^{\prime }=\sqrt{\tilde{p}_{9}^{\prime }}%
Y^{2}\text{, }A_{10}^{\prime }=\sqrt{\tilde{p}_{10}^{\prime }}Y^{3}\text{, }
\notag \\
&&  \notag \\
A_{11}^{\prime } &=&\sqrt{\tilde{p}_{11}^{\prime }}Y^{4}\text{, }%
A_{12}^{\prime }=\sqrt{\tilde{p}_{12}^{\prime }}Y^{5}\text{, }A_{13}^{\prime
}=\sqrt{\tilde{p}_{13}^{\prime }}Y^{6}\text{, }A_{14}^{\prime }=\sqrt{\tilde{%
p}_{14}^{\prime }}Y^{7}\text{, }A_{15}^{\prime }=\sqrt{\tilde{p}%
_{15}^{\prime }}Z^{1}\text{, }  \notag \\
&&  \notag \\
A_{16}^{\prime } &=&\sqrt{\tilde{p}_{16}^{\prime }}Z^{2}\text{, }%
A_{17}^{\prime }=\sqrt{\tilde{p}_{17}^{\prime }}Z^{3}\text{, }A_{18}^{\prime
}=\sqrt{\tilde{p}_{18}^{\prime }}Z^{4}\text{, }A_{19}^{\prime }=\sqrt{\tilde{%
p}_{19}^{\prime }}Z^{5}\text{, }A_{20}^{\prime }=\sqrt{\tilde{p}%
_{20}^{\prime }}Z^{6}\text{, }A_{21}^{\prime }=\sqrt{\tilde{p}_{21}^{\prime }%
}Z^{7}\text{.}  \label{w1}
\end{eqnarray}%
Finally, the $42$ weight-$2$ (correctable) error operators are,%
\begin{eqnarray}
A_{22}^{\prime } &=&\sqrt{\tilde{p}_{22}^{\prime }}X^{1}Z^{2}\text{, }%
A_{23}^{\prime }=\sqrt{\tilde{p}_{23}^{\prime }}X^{1}Z^{3}\text{, }%
A_{24}^{\prime }=\sqrt{\tilde{p}_{24}^{\prime }}X^{1}Z^{4}\text{, }%
A_{25}^{\prime }=\sqrt{\tilde{p}_{25}^{\prime }}X^{1}Z^{5}\text{, }%
A_{26}^{\prime }=\sqrt{\tilde{p}_{26}^{\prime }}X^{1}Z^{6}\text{,}  \notag \\
&&  \notag \\
\text{ }A_{27}^{\prime } &=&\sqrt{\tilde{p}_{27}^{\prime }}X^{1}Z^{7}\text{, 
}A_{28}^{\prime }=\sqrt{\tilde{p}_{28}^{\prime }}Z^{1}X^{2}\text{, }%
A_{29}^{\prime }=\sqrt{\tilde{p}_{29}^{\prime }}X^{2}Z^{3}\text{, }%
A_{30}^{\prime }=\sqrt{\tilde{p}_{30}^{\prime }}X^{2}Z^{4}\text{, }%
A_{31}^{\prime }=\sqrt{\tilde{p}_{31}^{\prime }}X^{2}Z^{5}\text{, }  \notag
\\
&&  \notag \\
A_{32}^{\prime } &=&\sqrt{\tilde{p}_{32}^{\prime }}X^{2}Z^{6}\text{, }%
A_{33}^{\prime }=\sqrt{\tilde{p}_{33}^{\prime }}X^{2}Z^{7}\text{, }%
A_{34}^{\prime }=\sqrt{\tilde{p}_{34}^{\prime }}Z^{1}X^{3}\text{, }%
A_{35}^{\prime }=\sqrt{\tilde{p}_{35}^{\prime }}Z^{2}X^{3}\text{, }%
A_{36}^{\prime }=\sqrt{\tilde{p}_{36}^{\prime }}X^{3}Z^{4}\text{,}  \notag \\
&&  \notag \\
\text{ }A_{37}^{\prime } &=&\sqrt{\tilde{p}_{37}^{\prime }}X^{3}Z^{5}\text{, 
}A_{38}^{\prime }=\sqrt{\tilde{p}_{38}^{\prime }}X^{3}Z^{6}\text{, }%
A_{39}^{\prime }=\sqrt{\tilde{p}_{39}^{\prime }}X^{3}Z^{7}\text{, }%
A_{40}^{\prime }=\sqrt{\tilde{p}_{40}^{\prime }}Z^{1}X^{4}\text{, }%
A_{41}^{\prime }=\sqrt{\tilde{p}_{41}^{\prime }}Z^{2}X^{4}\text{, }  \notag
\\
&&  \notag \\
A_{42}^{\prime } &=&\sqrt{\tilde{p}_{42}^{\prime }}Z^{3}X^{4}\text{, }%
A_{43}^{\prime }=\sqrt{\tilde{p}_{43}^{\prime }}X^{4}Z^{5}\text{, }%
A_{44}^{\prime }=\sqrt{\tilde{p}_{44}^{\prime }}X^{4}Z^{6}\text{, }%
A_{45}^{\prime }=\sqrt{\tilde{p}_{45}^{\prime }}X^{4}Z^{7}\text{, }%
A_{46}^{\prime }=\sqrt{\tilde{p}_{46}^{\prime }}Z^{1}X^{5}\text{, }
\end{eqnarray}%
and,%
\begin{eqnarray}
A_{47}^{\prime } &=&\sqrt{\tilde{p}_{47}^{\prime }}Z^{2}X^{5}\text{, }%
A_{48}^{\prime }=\sqrt{\tilde{p}_{48}^{\prime }}Z^{3}X^{5}\text{, }%
A_{49}^{\prime }=\sqrt{\tilde{p}_{49}^{\prime }}Z^{4}X^{5}\text{, }%
A_{50}^{\prime }=\sqrt{\tilde{p}_{50}^{\prime }}X^{5}Z^{6}\text{, }%
A_{51}^{\prime }=\sqrt{\tilde{p}_{51}^{\prime }}X^{5}Z^{7}\text{, }  \notag
\\
&&  \notag \\
A_{52}^{\prime } &=&\sqrt{\tilde{p}_{52}^{\prime }}Z^{1}X^{6}\text{, }%
A_{53}^{\prime }=\sqrt{\tilde{p}_{53}^{\prime }}Z^{2}X^{6}\text{, }%
A_{54}^{\prime }=\sqrt{\tilde{p}_{54}^{\prime }}Z^{3}X^{6}\text{, }%
A_{55}^{\prime }=\sqrt{\tilde{p}_{55}^{\prime }}Z^{4}X^{6}\text{, }%
A_{56}^{\prime }=\sqrt{\tilde{p}_{56}^{\prime }}Z^{5}X^{6}\text{, }  \notag
\\
&&  \notag \\
A_{57}^{\prime } &=&\sqrt{\tilde{p}_{57}^{\prime }}X^{6}Z^{7}\text{, }%
A_{58}^{\prime }=\sqrt{\tilde{p}_{58}^{\prime }}Z^{1}X^{7}\text{, }%
A_{59}^{\prime }=\sqrt{\tilde{p}_{59}^{\prime }}Z^{2}X^{7}\text{, }%
A_{60}^{\prime }=\sqrt{\tilde{p}_{60}^{\prime }}Z^{3}X^{7}\text{, }%
A_{61}^{\prime }=\sqrt{\tilde{p}_{61}^{\prime }}Z^{4}X^{7}\text{,}  \notag \\
&&  \notag \\
\text{ }A_{62}^{\prime } &=&\sqrt{\tilde{p}_{62}^{\prime }}Z^{5}X^{7}\text{, 
}A_{63}^{\prime }=\sqrt{\tilde{p}_{63}^{\prime }}Z^{6}X^{7}\text{.}
\end{eqnarray}

\emph{Detectable Errors }(Set-$1$).\emph{\ }For the sake of completeness, we
show in an explicit way that these $64$ errors are detectable. Considering $%
k\in \left\{ 0\text{, }1\text{,..., }21\right\} $, the error syndrome of
weight-$0$ and weight-$1$ error operators is given by,%
\begin{eqnarray}
S\left( I\right) &=&000000\text{, }S\left( X^{1}\right) =111000\text{, }%
S\left( X^{2}\right) =110000\text{, }S\left( X^{3}\right) =101000\text{, }%
S\left( X^{4}\right) =100000\text{,}  \notag \\
&&  \notag \\
\text{ }S\left( X^{5}\right) &=&011000\text{, }S\left( X^{6}\right) =010000%
\text{, }S\left( X^{7}\right) =001000\text{, }S\left( Y^{1}\right) =111111%
\text{, }S\left( Y^{2}\right) =110110\text{,}  \notag \\
&&  \notag \\
S\left( Y^{3}\right) &=&101101\text{, }S\left( Y^{4}\right) =100100\text{, }%
S\left( Y^{5}\right) =011011\text{, }S\left( Y^{6}\right) =010010\text{, }%
S\left( Y^{7}\right) =001001\text{,}  \notag \\
&&  \notag \\
S\left( Z^{1}\right) &=&000111\text{, }S\left( Z^{2}\right) =000110\text{, }%
S\left( Z^{3}\right) =000101\text{, }S\left( Z^{4}\right) =000100\text{, }%
S\left( Z^{5}\right) =000011\text{,}  \notag \\
&& \\
S\left( Z^{6}\right) &=&000010\text{, }S\left( Z^{7}\right) =000001\text{.} 
\notag
\end{eqnarray}%
Instead, for $k\in \left\{ 22\text{,..., }63\right\} $ the error syndrome of
weight-$2$ error operators is given by,%
\begin{eqnarray}
S\left( X^{1}Z^{2}\right) &=&111110\text{, }S\left( X^{1}Z^{3}\right) =111101%
\text{, }S\left( X^{1}Z^{4}\right) =111100\text{, }S\left( X^{1}Z^{5}\right)
=111011\text{, }S\left( X^{1}Z^{6}\right) =111010\text{,}  \notag \\
&&  \notag \\
\text{ }S\left( X^{1}Z^{7}\right) &=&111001\text{, }S\left(
Z^{1}X^{2}\right) =110111\text{, }S\left( X^{2}Z^{3}\right) =110101\text{, }%
S\left( X^{2}Z^{4}\right) =110100\text{, }S\left( X^{2}Z^{5}\right) =110011%
\text{, }  \notag \\
&&  \notag \\
S\left( X^{2}Z^{6}\right) &=&110010\text{, }S\left( X^{2}Z^{7}\right) =110001%
\text{, }S\left( Z^{1}X^{3}\right) =101111\text{, }S\left( Z^{2}X^{3}\right)
=101110\text{, }S\left( X^{3}Z^{4}\right) =101100\text{,}  \notag \\
&&  \notag \\
\text{ }S\left( X^{3}Z^{5}\right) &=&101011\text{, }S\left(
X^{3}Z^{6}\right) =101010\text{, }S\left( X^{3}Z^{7}\right) =101001\text{, }%
S\left( Z^{1}X^{4}\right) =100111\text{, }S\left( Z^{2}X^{4}\right) =100110%
\text{, }  \notag \\
&&  \notag \\
S\left( Z^{3}X^{4}\right) &=&100101\text{, }S\left( X^{4}Z^{5}\right) =100011%
\text{, }S\left( X^{4}Z^{6}\right) =100010\text{, }S\left( X^{4}Z^{7}\right)
=100001\text{, }S\left( Z^{1}X^{5}\right) =011111\text{, }
\end{eqnarray}%
and,%
\begin{eqnarray}
S\left( Z^{2}X^{5}\right) &=&011110\text{, }S\left( Z^{3}X^{5}\right) =011101%
\text{, }S\left( Z^{4}X^{5}\right) =011100\text{, }S\left( X^{5}Z^{6}\right)
=011010\text{, }S\left( X^{5}Z^{7}\right) =011001\text{, }  \notag \\
&&  \notag \\
S\left( Z^{1}X^{6}\right) &=&010111\text{, }S\left( Z^{2}X^{6}\right) =010110%
\text{, }S\left( Z^{3}X^{6}\right) =010101\text{, }S\left( Z^{4}X^{6}\right)
=010100\text{, }S\left( Z^{5}X^{6}\right) =010011\text{, }  \notag \\
&&  \notag \\
S\left( X^{6}Z^{7}\right) &=&010001\text{, }S\left( Z^{1}X^{7}\right) =001111%
\text{, }S\left( Z^{2}X^{7}\right) =001110\text{, }S\left( Z^{3}X^{7}\right)
=001101\text{, }S\left( Z^{4}X^{7}\right) =001100\text{,}  \notag \\
&&  \notag \\
\text{ }S\left( Z^{5}X^{7}\right) &=&001011\text{, }S\left(
Z^{6}X^{7}\right) =001010\text{.}
\end{eqnarray}%
Since these errors have non-vanishing error syndromes, they are detectable.
As a side remark, we point out that following the above-mentioned line of
reasoning, it can be shown that $S\left( A_{l}^{\prime \dagger
}A_{k}^{\prime }\right) \neq 0$, with $l$, $k\in \left\{ 0\text{, }1\text{%
,..., }63\right\} $.

\emph{Entanglement Fidelity }(Set-$1$). The explicit expression for $%
\mathcal{F}_{\text{Set-}1}^{\left[ \left[ 7,1,3\right] \right] }\left( \mu 
\text{, }p\right) $ in (\ref{S1}) is given by,%
\begin{eqnarray}
\mathcal{F}_{\text{Set-}1}^{\left[ \left[ 7,1,3\right] \right] }\left( \mu 
\text{, }p\right) &=&\mu ^{6}\left( \frac{4}{3}p^{7}-p^{6}-\frac{5}{3}p^{5}+%
\frac{4}{3}p^{4}\right) +\mu ^{5}\left( -8p^{7}+\frac{50}{3}p^{6}-\frac{22}{3%
}p^{5}-4p^{4}+\frac{8}{3}p^{3}\right) +  \notag \\
&&  \notag \\
&&\mu ^{4}\left( 20p^{7}-\frac{205}{3}p^{6}+\frac{250}{3}p^{5}-41p^{4}%
\allowbreak +\frac{14}{3}p^{3}+\frac{4}{3}p^{2}\right) +  \notag \\
&&  \notag \\
&&\mu ^{3}\left( -\frac{80}{3}p^{7}+\allowbreak \frac{380}{3}p^{6}-\frac{680%
}{3}p^{5}+192p^{4}-\frac{232}{3}p^{3}+12p^{2}\right) +  \notag \\
&&  \notag \\
&&\mu ^{2}\left( 20p^{7}\allowbreak -\frac{365}{3}p^{6}+\frac{835}{3}%
p^{5}\allowbreak -310p^{4}+\frac{530}{3}p^{3}-\frac{145}{3}p^{2}+5p\right) +
\notag \\
&&  \notag \\
&&\mu \left( -8p^{7}+\frac{178}{3}p^{6}-\frac{490}{3}p^{5}+220p^{4}-\frac{460%
}{3}\allowbreak p^{3}+\allowbreak \frac{154}{3}p^{2}-6p\right) +  \notag \\
&&  \notag \\
&&\left( \frac{4}{3}p^{7}-\frac{35}{3}p^{6}+\frac{112}{3}p^{5}-\frac{175}{3}%
p^{4}+\frac{140}{3}p^{3}-\frac{49}{3}p^{2}+1\right) \text{.}
\end{eqnarray}

\emph{Error Operators }(Set-$2$). The sets of weight-$0$ and weight-$1$
correctable errors are given in (\ref{w0}) and (\ref{w1}), respectively. The
chosen set of correctable weight-$2$ error operators is,%
\begin{eqnarray}
A_{22}^{^{\prime \prime }} &=&\sqrt{\tilde{p}_{22}^{^{\prime \prime }}}%
Z^{1}Z^{2}\text{, }A_{23}^{^{\prime \prime }}=\sqrt{\tilde{p}_{23}^{^{\prime
\prime }}}Z^{1}Z^{3}\text{, }A_{24}^{^{\prime \prime }}=\sqrt{\tilde{p}%
_{24}^{^{\prime \prime }}}Z^{1}Z^{4}\text{, }A_{25}^{^{\prime \prime }}=%
\sqrt{\tilde{p}_{25}^{^{\prime \prime }}}Z^{1}Z^{5}\text{, }A_{26}^{^{\prime
\prime }}=\sqrt{\tilde{p}_{26}^{^{\prime \prime }}}Z^{1}Z^{6}\text{,}  \notag
\\
&&  \notag \\
\text{ }A_{27}^{^{\prime \prime }} &=&\sqrt{\tilde{p}_{27}^{^{\prime \prime
}}}Z^{1}Z^{7}\text{, }A_{28}^{^{\prime \prime }}=\sqrt{\tilde{p}%
_{28}^{^{\prime \prime }}}Z^{2}Z^{3}\text{, }A_{29}^{^{\prime \prime }}=%
\sqrt{\tilde{p}_{29}^{^{\prime \prime }}}Z^{2}Z^{4}\text{, }A_{30}^{^{\prime
\prime }}=\sqrt{\tilde{p}_{30}^{\prime \prime }}Z^{2}Z^{5}\text{, }%
A_{31}^{^{\prime \prime }}=\sqrt{\tilde{p}_{31}^{^{\prime \prime }}}%
Z^{2}Z^{6}\text{, }  \notag \\
&&  \notag \\
A_{32}^{^{\prime \prime }} &=&\sqrt{\tilde{p}_{32}^{^{\prime \prime }}}%
Z^{2}Z^{7}\text{, }A_{33}^{^{\prime \prime }}=\sqrt{\tilde{p}_{33}^{^{\prime
\prime }}}Z^{3}Z^{4}\text{, }A_{34}^{^{\prime \prime }}=\sqrt{\tilde{p}%
_{34}^{^{\prime \prime }}}Z^{3}Z^{5}\text{, }A_{35}^{^{\prime \prime }}=%
\sqrt{\tilde{p}_{35}^{^{\prime \prime }}}Z^{3}Z^{6}\text{, }A_{36}^{^{\prime
\prime }}=\sqrt{\tilde{p}_{36}^{^{\prime \prime }}}Z^{3}Z^{7}\text{,}  \notag
\\
&&  \notag \\
\text{ }A_{37}^{^{\prime \prime }} &=&\sqrt{\tilde{p}_{37}^{^{\prime \prime
}}}Z^{4}Z^{5}\text{, }A_{38}^{^{\prime \prime }}=\sqrt{\tilde{p}%
_{38}^{^{\prime \prime }}}Z^{4}Z^{6}\text{, }A_{39}^{^{\prime \prime }}=%
\sqrt{\tilde{p}_{39}^{^{\prime \prime }}}Z^{4}Z^{7}\text{, }A_{40}^{^{\prime
\prime }}=\sqrt{\tilde{p}_{40}^{^{\prime \prime }}}Z^{5}Z^{6}\text{, }%
A_{41}^{^{\prime \prime }}=\sqrt{\tilde{p}_{41}^{^{\prime \prime }}}%
Z^{5}Z^{7}\text{, }  \notag \\
&&  \notag \\
A_{42}^{^{\prime \prime }} &=&\sqrt{\tilde{p}_{42}^{^{\prime \prime }}}%
Z^{6}Z^{7}\text{, }
\end{eqnarray}%
and,%
\begin{eqnarray}
A_{43}^{^{\prime \prime }} &=&\sqrt{\tilde{p}_{43}^{^{\prime \prime }}}%
Z^{1}X^{2}\text{, }A_{44}^{^{\prime \prime }}=\sqrt{\tilde{p}_{44}^{^{\prime
\prime }}}Z^{1}X^{3}\text{, }A_{45}^{^{\prime \prime }}=\sqrt{\tilde{p}%
_{45}^{^{\prime \prime }}}Z^{1}X^{4}\text{, }A_{46}^{^{\prime \prime }}=%
\sqrt{\tilde{p}_{46}^{^{\prime \prime }}}Z^{1}X^{5}\text{, }  \notag \\
&&  \notag \\
A_{47}^{^{\prime \prime }} &=&\sqrt{\tilde{p}_{47}^{^{\prime \prime }}}%
Z^{1}X^{6}\text{, }A_{48}^{^{\prime \prime }}=\sqrt{\tilde{p}_{48}^{^{\prime
\prime }}}Z^{1}X^{7}\text{, }A_{49}^{^{\prime \prime }}=\sqrt{\tilde{p}%
_{49}^{^{\prime \prime }}}Z^{2}X^{3}\text{, }A_{50}^{^{\prime \prime }}=%
\sqrt{\tilde{p}_{50}^{^{\prime \prime }}}Z^{2}X^{4}\text{, }A_{51}^{^{\prime
\prime }}=\sqrt{\tilde{p}_{51}^{^{\prime \prime }}}Z^{2}X^{5}\text{, } 
\notag \\
&&  \notag \\
A_{52}^{^{\prime \prime }} &=&\sqrt{\tilde{p}_{52}^{^{\prime \prime }}}%
Z^{2}X^{6}\text{, }A_{53}^{^{\prime \prime }}=\sqrt{\tilde{p}_{53}^{^{\prime
\prime }}}Z^{2}X^{7}\text{, }A_{54}^{^{\prime \prime }}=\sqrt{\tilde{p}%
_{54}^{^{\prime \prime }}}Z^{3}X^{4}\text{, }A_{55}^{^{\prime \prime }}=%
\sqrt{\tilde{p}_{55}^{^{\prime \prime }}}Z^{3}X^{5}\text{, }A_{56}^{^{\prime
\prime }}=\sqrt{\tilde{p}_{56}^{^{\prime \prime }}}Z^{3}X^{6}\text{, } 
\notag \\
&&  \notag \\
A_{57}^{^{\prime \prime }} &=&\sqrt{\tilde{p}_{57}^{^{\prime \prime }}}%
Z^{3}X^{7}\text{, }A_{58}^{^{\prime \prime }}=\sqrt{\tilde{p}_{58}^{^{\prime
\prime }}}Z^{4}X^{5}\text{, }A_{59}^{^{\prime \prime }}=\sqrt{\tilde{p}%
_{59}^{^{\prime \prime }}}Z^{4}X^{6}\text{, }A_{60}^{^{\prime \prime }}=%
\sqrt{\tilde{p}_{60}^{^{\prime \prime }}}Z^{4}X^{7}\text{, }A_{61}^{^{\prime
\prime }}=\sqrt{\tilde{p}_{61}^{^{\prime \prime }}}Z^{5}X^{6}\text{,}  \notag
\\
&&  \notag \\
\text{ }A_{62}^{^{\prime \prime }} &=&\sqrt{\tilde{p}_{62}^{^{\prime \prime
}}}Z^{5}X^{7}\text{, }A_{63}^{^{\prime \prime }}=\sqrt{\tilde{p}%
_{63}^{^{\prime \prime }}}Z^{6}X^{7}\text{.}
\end{eqnarray}

\emph{Entanglement Fidelity} (Set-$2$). The explicit expression for $%
\mathcal{F}_{\text{Set-}2}^{\left[ \left[ 7,1,3\right] \right] }\left( \mu 
\text{, }p\right) $ in (\ref{S2}) is given by,%
\begin{eqnarray}
\mathcal{F}_{\text{Set-}2}^{\left[ \left[ 7\text{, }1\text{, }3\right] %
\right] }\left( \mu \text{, }p\right) &=&\mu ^{6}\left( \frac{4}{3}%
p^{7}+p^{6}-5p^{5}+\frac{8}{3}p^{4}\right) +\mu ^{5}\left( -8p^{7}+\frac{20}{%
3}p^{6}+16p^{5}-\frac{64}{3}p^{4}+\frac{20}{3}p^{3}\right) +  \notag \\
&&  \notag \\
&&\mu ^{4}\left( \allowbreak 20p^{7}-\frac{145}{3}p^{6}+\frac{70}{3}%
p^{5}+23p^{4}-\frac{70}{3}p^{3}+\frac{16}{3}p^{2}\allowbreak \right) + 
\notag \\
&&  \notag \\
&&\mu ^{3}\left( -\frac{80}{3}p^{7}+\frac{320}{3}p^{6}-\frac{460}{3}p^{5}+%
\frac{272}{3}p^{4}-\frac{40}{3}p^{3}-\frac{16}{3}\allowbreak p^{2}+\frac{4}{3%
}p\right) +  \notag \\
&&  \notag \\
&&\mu ^{2}\left( \allowbreak 20p^{7}-\frac{335}{3}p^{6}+235p^{5}-\frac{710}{3%
}p^{4}+\frac{350}{3}p^{3}-25\allowbreak p^{2}+\frac{5}{3}p\right) +  \notag
\\
&&  \notag \\
&&\mu \left( -8p^{7}+\frac{172}{3}p^{6}-\frac{460}{3}p^{5}+200p^{4}-\frac{400%
}{3}p^{3}+\frac{124}{3}\allowbreak p^{2}-4p\right) +  \notag \\
&&  \notag \\
&&\left( \frac{4}{3}p^{7}-\frac{35}{3}p^{6}+\frac{112}{3}p^{5}-\frac{175}{3}%
p^{4}+\frac{140}{3}p^{3}-\frac{49}{3}p^{2}+1\right) \text{.}
\end{eqnarray}

\section{Asymmetries and Correlations}

\emph{Entanglement Fidelity}. Substituting (\ref{x}) in (\ref{xx}) the
explicit expression for $\mathcal{F}_{\text{asymmetric}}^{\left[ \left[ 7,1,3%
\right] \right] }\left( \mu \text{, }p\right) $ becomes,%
\begin{equation}
\mathcal{F}_{\text{asymmetric}}^{\left[ \left[ 7,1,3\right] \right] }\left(
\mu \text{, }p\right) =\mathcal{A}_{6}\left( \mu \text{, }p\right) +\mathcal{%
A}_{5}\left( \mu \text{, }p\right) +\mathcal{A}_{4}\left( \mu \text{, }%
p\right) +\mathcal{A}_{3}\left( \mu \text{, }p\right) +\mathcal{A}_{2}\left(
\mu \text{, }p\right) +\mathcal{A}_{1}\left( \mu \text{, }p\right) +\mathcal{%
A}_{0}\left( \mu \text{, }p\right) \text{.}  \label{asymG}
\end{equation}%
The quantities $\mathcal{A}_{6}\left( \mu \text{, }p\right) $, $\mathcal{A}%
_{5}\left( \mu \text{, }p\right) $, $\mathcal{A}_{4}\left( \mu \text{, }%
p\right) $ and $\mathcal{A}_{3}\left( \mu \text{, }p\right) $ are given by,%
\begin{eqnarray}
\mathcal{A}_{6}\left( \mu \text{, }p\right) &=&\text{ }\mu ^{6}\left[ \left(
6p^{7}-11p^{6}+\allowbreak 5p^{5}\right) +\alpha _{Z}\left(
6p^{6}-10p^{5}+4p^{4}\right) +\left( \alpha _{Z}^{2}+\alpha _{X}\alpha
_{Z}\right) \left( -21p^{7}\allowbreak +45p^{6}-\allowbreak
30p^{5}+6p^{4}\right) \right] \text{,}  \notag \\
&&  \notag \\
&&  \notag \\
\mathcal{A}_{5}\left( \mu \text{, }p\right) &=&\text{ }\mu ^{5}\left[ 
\begin{array}{c}
\left( -36p^{7}+90p^{6}-74p^{5}+20p^{4}\right) +\alpha _{Z}\left(
-30p^{6}+70p^{5}-52p^{4}+12p^{3}\right) + \\ 
\\ 
+\left( \alpha _{Z}^{2}+\alpha _{X}\alpha _{Z}\right) \left(
126p^{7}\allowbreak -330p^{6}+300p^{5}-108p^{4}+12p^{3}\right)%
\end{array}%
\right] \text{,}  \notag \\
&&  \notag \\
&&  \notag \\
\mathcal{A}_{4}\left( \mu \text{, }p\right) &=&\text{ }\mu ^{4}\left[ 
\begin{array}{c}
\left( 90p^{7}-285\allowbreak p^{6}+330p^{5}-165p^{4}+30\allowbreak
p^{3}\right) +\alpha _{Z}\left( 60p^{6}\allowbreak
-180p^{5}+192p^{4}-84p^{3}+12p^{2}\allowbreak \right) + \\ 
\\ 
+\left( \alpha _{Z}^{2}+\alpha _{X}\allowbreak \alpha _{Z}\right) \left(
-315p^{7}+975p^{6}-\allowbreak 1110p^{5}+558p^{4}-114p^{3}+6p^{2}\right)%
\end{array}%
\right] \text{,}  \notag \\
&&  \notag \\
&&  \notag \\
\mathcal{A}_{3}\left( \mu \text{, }p\right) &=&\mu ^{3}\left[ 
\begin{array}{c}
\left( -120p^{7}+460p^{6}-680p^{5}+\allowbreak 480p^{4}-160p^{3}+\allowbreak
20p^{2}\right) + \\ 
\\ 
+\alpha _{Z}\left( -60p^{6}+220p^{5}-304p^{4}+192p^{3}-52p^{2}+4\allowbreak
p\right) + \\ 
\\ 
+\left( \alpha _{Z}^{2}+\alpha _{X}\alpha _{Z}\right) \left(
420p^{7}-1500p^{6}+2040\allowbreak p^{5}-1296p^{4}+372p^{3}-36p^{2}\right)%
\end{array}%
\right] \text{,}
\end{eqnarray}%
while $\mathcal{A}_{2}\left( \mu \text{, }p\right) $, $\mathcal{A}_{1}\left(
\mu \text{, }p\right) $ and $\mathcal{A}_{0}\left( \mu \text{, }p\right) $
are, 
\begin{eqnarray}
\mathcal{A}_{2}\left( \mu \text{, }p\right) &=&\text{ }\mu ^{2}\left[ 
\begin{array}{c}
\left( 90\allowbreak p^{7}-405p^{6}+725\allowbreak
p^{5}-650p^{4}+300p^{3}-65p^{2}+5p\right) + \\ 
\\ 
+\alpha _{Z}\left( \allowbreak 30p^{6}-130p^{5}+\allowbreak
220p^{4}-180p^{3}+70\allowbreak p^{2}-10p\right) + \\ 
\\ 
+\left( \alpha _{Z}^{2}+\alpha _{X}\alpha _{Z}\right) \left( \allowbreak
\allowbreak -315p^{7}+1275p^{6}-2010\allowbreak
p^{5}+1530p^{4}-555p^{3}+75p^{2}\right)%
\end{array}%
\right] \text{,}  \notag \\
&&  \notag \\
&&  \notag \\
\mathcal{A}_{1}\left( \mu \text{, }p\right) &=&\text{ }\mu \left[ 
\begin{array}{c}
\left( -36p^{7}+186p^{6}-390p^{5}+420p^{4}-240p^{3}+66p^{2}-6\allowbreak
p\right) + \\ 
\\ 
+\alpha _{Z}\left( -6\allowbreak
p^{6}+30p^{5}-60p^{4}+60p^{3}-30p^{2}+6p\right) + \\ 
\\ 
+\left( \alpha _{Z}^{2}+\alpha _{X}\alpha _{Z}\right) \left( \allowbreak
126p^{7}-570p^{6}+\allowbreak 1020\allowbreak
p^{5}-900p^{4}+390p^{3}-66p^{2}\right)%
\end{array}%
\right] \text{,}  \notag \\
&&  \notag \\
&&  \notag \\
\mathcal{A}_{0}\left( \mu \text{, }p\right) &=&\text{ }\left( 1-p\right)
^{7}+7p\left( 1-p\right) ^{6}+21p^{2}\left( 1-p\right) ^{5}\left[ \alpha
_{Z}^{2}+\alpha _{X}\alpha _{Z}\right] \text{.}
\end{eqnarray}

\end{document}